\documentclass[%
 reprint,
 superscriptaddress,
 nofootinbib,
 amsmath,amssymb,
 aps,
 prd,
]{revtex4-2}

\usepackage{bm}
\usepackage{xcolor}
\usepackage{dcolumn}
\usepackage{multirow}
\usepackage{amsfonts}
\usepackage{graphicx}
\usepackage{slashed}
\usepackage{bbm}
\usepackage{makecell}
\usepackage[normalem]{ulem}

\usepackage{hyperref}
\usepackage{orcidlink}

\newenvironment{mymathbox}
{\par\smallskip\centering\begin{lrbox}{0}%
\begin{minipage}[c]{0.8\textwidth}}
{\end{minipage}\end{lrbox}%
\framebox[0.9\textwidth]{\usebox{0}}%
\par\medskip
\ignorespacesafterend}
\newcommand{\bb}{\begin{mymathbox}}
\newcommand{\eb}{\end{mymathbox}}

\newcommand{\be}{\begin{equation}}
\newcommand{\ee}{\end{equation}}
\newcommand{\ba}{\begin{eqnarray}}
\newcommand{\ea}{\end{eqnarray}}

\newcommand{\np}{{\bf      p}}

\newcommand{\npsi}{{\bf \npsi}}

\newcommand{\Psib}{\overline{\Psi}}

\newcommand{\bma}{\begin{pmatrix}}
\newcommand{\ema}{\end{pmatrix}}

\begin{document}

\title{Two-body current and axial form factor effects in charged-current quasielastic neutrino-nucleus scattering within the NEUT event generator}

\author{T. Franco-Munoz\,\orcidlink{0009-0000-3871-4752}}
\email{tania.francomunoz@ugent.be}
\affiliation{Department of Physics and Astronomy, Ghent University, B-9000 Gent, Belgium}
\author{J.~McKean\,\orcidlink{0009-0005-6100-6195}\,}
 \email{mckean.jake.42u@st.kyoto-u.ac.jp}
 \affiliation{Kyoto University, Department of Physics, Kyoto, Japan}
\author{J.~Garc\'ia-Marcos\,\orcidlink{0009-0003-2753-1864}}
\affiliation{Department of Physics and Astronomy, Ghent University, B-9000 Gent, Belgium}
\affiliation{Grupo de F\'isica Nuclear, Departamento de Estructura de la Materia, F\'isica T\'ermica y Electr\'onica, Facultad de Ciencias F\'isicas, Universidad Complutense de Madrid and IPARCOS, CEI Moncloa, Madrid 28040, Spain}
\author{M.~Hooft\,\orcidlink{0000-0002-6368-4820}}
\affiliation{Department of Physics and Astronomy, Ghent University, B-9000 Gent, Belgium}
\author{R.~Gonz\'alez-Jim\'enez\,\orcidlink{0000-0002-0492-0619}}
\affiliation{Departamento de F\'isica At\'omica, Molecular y Nuclear, Universidad de Sevilla, 41080 Sevilla, Spain}
\author{N.~Jachowicz\,\orcidlink{0000-0003-1168-0745}}
\affiliation{Department of Physics and Astronomy, Ghent University, B-9000 Gent, Belgium}
\author{J.~M.~Ud\'ias\,\orcidlink{0000-0003-3714-764X}}
\affiliation{Grupo de F\'isica Nuclear, Departamento de Estructura de la Materia, F\'isica T\'ermica y Electr\'onica, Facultad de Ciencias F\'isicas, Universidad Complutense de Madrid and IPARCOS, CEI Moncloa, Madrid 28040, Spain}

\date{\today}

\begin{abstract}
We present a charged-current quasielastic neutrino-nucleus scattering model based on an unfactorized representation of the spectral function, employing relativistic momentum distributions for bound nucleons and the relativistic distorted-wave impulse approximation with an energy-dependent relativistic potential to describe the scattered nucleon. The model incorporates two-body meson-exchange currents contributing to one-particle-one-hole final states and tests several axial form factor parametrizations, including recent LQCD and MINERvA fits. It is implemented in the NEUT event generator and benchmarked against T2K and MINERvA $\nu_\mu$-$^{12}$C CC0$\pi$ measurements. We find that two-body meson-exchange currents lead to a sizeable increase of the total cross section, arising from an enhancement of the transverse response, which is the dominant component in charged-current neutrino scattering. On the other hand, recent fits of the axial form factor predict larger values than the standard dipole form, yielding a systematic enhancement of the cross section. The LQCD+MINERvA parametrization tends to overestimate the data, while the MINERvA-only fit provides a more moderate increase. Overall, no single configuration consistently provides the best agreement with the different datasets. 
\end{abstract}

\maketitle

\section{Introduction}

Precise measurements of neutrino properties are among the highest priorities in fundamental particle physics, involving worldwide efforts through current experiments (e.g. T2K~\cite{T2K20}, NOvA~\cite{NOvA19} and MicroBooNE~\cite{MicroBooNE20}), together with recent measurements from MINERvA~\cite{MINERvA20}, as well as next-generation programs (e.g. Hyper-Kamiokande~\cite{HyperK15} and DUNE~\cite{DUNE16}). With these future programs, the field will enter the precision era, significantly reducing statistical uncertainties and enhancing the sensitivity to the details of neutrino-nucleus interactions. In this context, however, the wealth of electron scattering data available, together with the growing collection of measurements from accelerator-based neutrino experiments, have revealed limitations in the lepton-nucleus cross section models commonly employed in neutrino event generators~\cite{Abe21,Khachatryan21,Abratenko23,Kleykamp23}. As a result, neutrino interaction systematics are expected to become the dominant source of uncertainty~\cite{Alvarez-Ruso18}, making it essential to improve our ability to describe neutrino-nucleus cross sections with high accuracy for the success of future neutrino oscillation experiments.

Accelerator-based experiments use broad-band neutrino fluxes. Hence, neutrino-nucleus interactions involve multiple reaction mechanisms,  each dominating at different incident neutrino energies. This requires robust models that properly describe the variety of mechanisms occurring over the whole experimental range. Among these, charged-current quasielastic (CCQE) scattering provides a significant contribution to the signal in current and future oscillation experiments, being the dominant contribution to the overall neutrino event rate in the low-energy experiments T2K and MicroBooNE, and a major contribution in MINERvA, NOvA and DUNE.

Motivated by this situation, in this work we focus on the CCQE interaction in order to improve the description of this channel for neutrino Monte Carlo simulations. To this end, our fully relativistic and quantum mechanical model from Refs.~\cite{Franco-Munoz23,Franco-Munoz25}, initially developed and validated for the case of electromagnetic interactions, is here extended to the charged-current neutrino sector. CCQE neutrino-nucleus scattering is then studied using an unfactorized representation of the spectral function (SF), where the momentum distributions of bound nucleons are obtained from a relativistic mean field model for finite nuclei, and the final-state nucleon is described as a scattering solution of the Dirac equation with relativistic potentials.  Our framework also includes the contribution from two-body meson-exchange currents (MEC) to a one-particle-one-hole (1p1h) final state. In the case of inclusive electromagnetic scattering, their impact has been observed to be significant only in the transverse channel, where the response is increased, while the longitudinal component remains largely unaffected. The good agreement with electron scattering experimental data supports the predicted enhancement of the transverse response by the two-body currents, highlighting their essential role for a complete and consistent description of the analogous neutrino-induced process.

The extension of the formalism to neutrino interactions introduces an axial component that is absent in electron scattering. The one-body current, which provides the dominant contribution to the CCQE channel, depends on the axial form factor; this is significantly less constrained than its vector counterpart, making it a leading source of systematic uncertainty.  Historically, the axial form factor has been described using a simple dipole parametrization. However, this functional form may not accurately capture the range of shapes compatible with theoretical constraints and existing experimental data~\cite{Bernard02,Bodek08}. In addition, recent results from lattice-QCD (LQCD) calculations \cite{Meyer26} and hydrogen-target data from the MINERvA collaboration \cite{MINERvA23,Meyer25} suggest that the axial form factor may be underestimated at large momentum transfer when using the conventional dipole form.

These developments have been integrated into the NEUT Monte Carlo event generator, where our unfactorized spectral function framework was first implemented at the one-body current level~\cite{McKean25}. Building upon that work, the present study extends the NEUT implementation by including two-body meson-exchange current contributions and by exploring the impact of different axial form factor parametrizations. In this way, we assess how these improvements affect neutrino-nucleus cross sections and the associated event kinematics within a framework directly applicable to oscillation analyses.

The paper is organized as follows. In Sec.~\ref{sec:model}, we present our theoretical model, including the description of the nuclear structure and dynamics together with the one- and two-body current contributions. In Sec.~\ref{sec:results}, we show our results for neutrino-nucleus cross sections and their comparison with experimental data from the T2K and MINERvA collaborations. Finally, Sec.~\ref{sec:conclusions} summarizes the main conclusions of this work.
\section{Theoretical Framework}\label{sec:model}

The modeling of the (anti)neutrino-nucleus CCQE scattering process is performed within the first-order Born approximation, in which a single $W$ boson is exchanged between the lepton and the nuclear system, coupling to a bound nucleon inside the target nucleus, and the leptons are described as free Dirac particles.

\begin{figure}[hbtp]
\includegraphics[width=\columnwidth,angle=0]{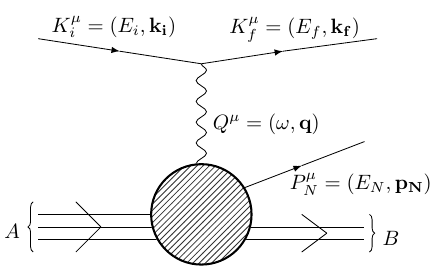}
\caption{Kinematics of the one-nucleon knock-out lepton-nucleus interaction. An initial lepton with four-momentum $K_i^\mu$ is scattered into a final state with $K_f^\mu$ via the exchange of a virtual boson carrying four-momentum $Q^\mu$. We work in the laboratory reference frame, where the target nucleus $A$ is at rest. The interaction leads to a one-particle-one-hole final state with an outgoing nucleon of four-momentum $P_N^\mu$ and a residual nuclear system $B$.}
\label{fig:process}
\end{figure}

The differential cross-section for the detection of a lepton in coincidence with a knocked-out nucleon which has interacted with the exchanged boson (see Fig.~\ref{fig:process}) is 
\begin{multline}
    \frac{d^{6}\sigma}{dk_f d\Omega_f d\Omega_N dE_m}= \\
    \sum_\kappa \rho_\kappa(E_m) \frac{G_F^2 \cos^2{\theta_c}}{2} \frac{K}{(2\pi)^5} L_{\mu\nu} H^{\mu\nu}_\kappa,
\label{eq:cross_section}
\end{multline}
where $\rho_\kappa(E_m)$ represents the missing energy~\footnote{By missing energy we refer to the part of the transferred energy that is converted into internal excitation of the residual nucleus.} profile for the nuclear state $\kappa$, and the cross section is decomposed in terms of leptonic and hadronic tensors, $L_{\mu\nu}$ and $H^{\mu\nu}$, allowing us to treat the lepton and hadron components separately. $G_F$ is the Fermi constant and $\theta_c$ the Cabibbo mixing angle. In addition, the $Q^2$ dependence of the intermediate $W$ boson propagator has been neglected as the energies involved in the process are significantly smaller than its mass. $K$ is a function that contains kinematical factors,
\begin{align}
K &=\dfrac{k_f^2}{E_i E_f} \frac{p_N E_N M_B}{M_A f_{rec}}, \\
f_{rec} &= \left| 1+ \frac{\omega p_N - q E_N \cos{\theta_N}}{M_A p_N}  \right|.
\end{align} 
where $\omega$ and $q$ denote the energy and momentum transfer, $M_A$ and $M_B$ are the masses of the target and residual system and the subscripts $i$ and $f$ denote the initial and final lepton variables, while the subscript $N$ refers to the outgoing nucleon.

The ground state of the target nucleus is modeled using an unfactorized representation of the spectral function, that is built from information extracted from $(e,e'p)$ measurements over a broad kinematic range~\cite{Udias93, Giusti11, Atkinson18, Egiyan06, Duer18}. These data, covering regions both below and above the two-nucleon emission threshold, has led to the development of semi-phenomenological spectral functions~\cite{Benhar94, Ankowski24},  that are composed of discrete and continuum components. In our work, we use this information to build an unfactorized calculation of the cross section which, were factorization employed, would yield virtually the same results as the semi-phenomenological spectral function in a factorized approach.

In these semi-phenomenological spectral functions, the discrete component originates from one-hole contributions to the final state at low and intermediate missing energies. These contributions are sharply peaked around the energies of the one-hole states in the target nucleus, i.e. the states obtained by removing one nucleon from the target nucleus, leaving the residual system in a discrete (not resonant) state. We use energies and, to some extent, widths taken from the missing-energy spectra where one-hole states are experimentally observed, rather than relying on mean-field predictions.  One-hole states can differ significantly from mean-field expectations, and they usually do. They often appear fragmented into several states and carry only a fraction (typically $1/3$ to $3/4$) of the mean-field strength. 
We could introduce here the whole terminology of quasi-particle and spectral amplitude, that is the right wording from many-body theory, and prevent any mention to the mean-field calculation. We only employ the momentum (or spatial) distribution from mean-field solutions to parametrize the $p$- (or $r$-)dependence of the quasi-particle function or the one-hole spectral amplitude. This is justified by the fact that both actual many-body calculations and $(e,e'p)$ experiments show that these mean-field distributions accurately reproduce the shape (though not the amplitude) of the many-body spectral amplitude or quasi-particle function. The continuum part arises from the correlated nature of nuclear states, which makes it possible to knock-out more than one nucleon in the final state, even when the exchanged boson interacts with only one nucleon. The missing energy would contain the kinetic energy of any additional knocked-out nucleons; in this regime, it becomes a continuum quantity that opens only above the two-nucleon threshold. The total probability to knock-out a nucleon is split between the discrete one-hole contributions and the continuum multi-nucleon final state outcome. Thus, in general and compared to the mean-field distribution of strength against missing energy, in an actual nucleus only a fraction of the total nucleon-knockout strength is observed below the two-nucleon threshold, while the remaining strength is shifted to higher missing energies. Within this framework, the index $\kappa$ labels the different contributions to the spectral function, accounting for both the discrete one-hole states and the correlated continuum component. The missing-energy profile, $\rho(E_m)$, is constructed to reproduce that of the semi-phenomenological spectral function in a form suitable for our unfactorized calculation. For $^{12}$C, we take the Rome SF~\cite{Benhar94, Benhar05} as a guide (see Refs.~\cite{Gonzalez-Jimenez22, Franco-Patino22, Franco-Munoz23} for details).

The leptonic component of the cross section can be readily obtained, leading to
\begin{multline}
    L_{\mu\nu} =  2 [K_{i,\mu} K_{f,\nu} + K_{i,\nu} K_{f,\mu} \\ - g_{\mu\nu} (K_i \cdot K_f ) 
    - i h \epsilon_{\mu\nu\alpha\beta} K_i^\alpha K_f^\beta ],
\end{multline}
with $h=-1(+1)$ for (anti)neutrinos and $\epsilon^{\mu\nu\alpha\beta}$ the Levi-Civita symbol ($\epsilon^{0123}=1$).

On the other hand, the hadronic tensor requires further study, as it contains all the information about the boson-nucleus interaction as well as the hadronic final-state interactions. The hadronic tensor for the interaction with a nucleon in the $\kappa$ state is defined as
\be
    H^{\mu\nu}_\kappa = \frac{1}{2j+1}\sum_{m_j}\sum_{s_N} J_\kappa^\mu (J_\kappa^\nu)^\dagger.
\ee
$J_\kappa^\mu$ is the hadronic current which, after integration over the degrees of freedom of the spectator nucleons, is written in terms of quasi-particle quantities and then computed from the summed contributions of single-particle-like matrix element calculations as \cite{Udias93}
\be
   J_\kappa^\mu=  \sqrt{V} \int{d\np}    \hspace{1.5mm} \Psib^{s_N}(\np_N,\np_N') \Gamma^{\mu}  \Psi_\kappa^{m_j}(\np). 
   \label{eq:hadronic-current_1b}
\ee
The momentum distribution of the bound nucleon, $\Psi_\kappa^{m_j}$, is obtained within the relativistic mean-field model of~\cite{Sharma93}, with $\np$ its momentum and $m_j$ the third-component of its total angular momentum $j$.  
The final-state nucleon, $\Psi^{s_N}$, is described within the relativistic distorted wave impulse approximation as a solution of the Dirac equation in the continuum using the energy-dependent relativistic  (ED-RMF) potential of ~\cite{Gonzalez-Jimenez19,Nikolakopoulos19,Gonzalez-Jimenez20}, which accounts for elastic final-state interactions with the residual system.  As it corresponds to a real potential, the total flux is conserved. The outgoing nucleon is characterized by its asymptotic momentum $\np_N$ and spin $s_N$, while $\np_N'$, given by ${\bf p}+{\bf q}$ from momentum conservation at the interaction vertex, denotes its momentum inside the nucleus.

\begin{figure}[hbtp]
\includegraphics[width=0.9\columnwidth,angle=0]{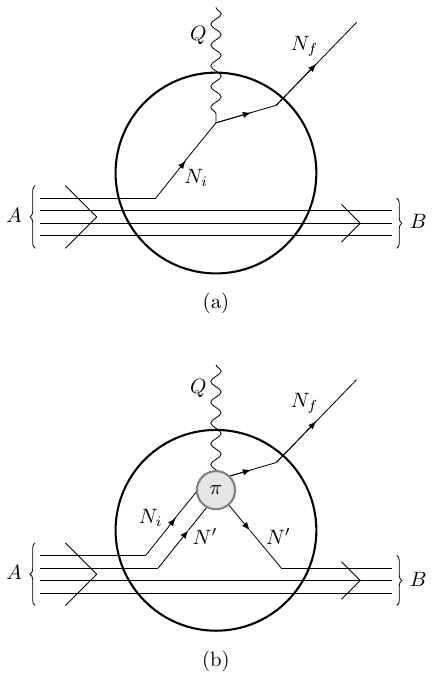}
\caption{Diagrammatic representation of the one- and two-body contributions to the lepton-nucleus interaction leading to a 1p1h final state. Panel (a) corresponds to the one-body mechanism, while panel (b) represents the two-body meson-exchange current contribution. The pion-exchange bubble accounts for the different two-body MEC diagrams shown in Figs.~\ref{fig:delta_diagrams} and \ref{fig:chpt_diagrams}.  $N_i$ and $N_f$ denote the initial and final nucleons, respectively, while $N'$ denotes the second bound nucleon  involved in the two-body interaction.}
\label{fig:operators}
\end{figure}

In this work, the hadronic current operator, $\Gamma^\mu$, includes the one-body term together with the contribution from two-body meson-exchange currents leading to a final 1p1h state,
\ba
   \Gamma^\mu= \Gamma_{1b}^\mu + \Gamma_{2b}^\mu .
\ea
These mechanisms are illustrated in Fig.~\ref{fig:operators}. In the former case, the lepton interacts with a single nucleon, which is subsequently knocked out from the nucleus (Fig.~\ref{fig:operators}(a)). Meanwhile, the two-body contribution accounts for those processes in which the initially struck nucleon interacts with another bound nucleon in the target nucleus through pion exchange, leading to the knock-out of one of the involved nucleons (Fig.~\ref{fig:operators}(b)).

\begin{figure}[hbtp]
\includegraphics[width=\columnwidth,angle=0]{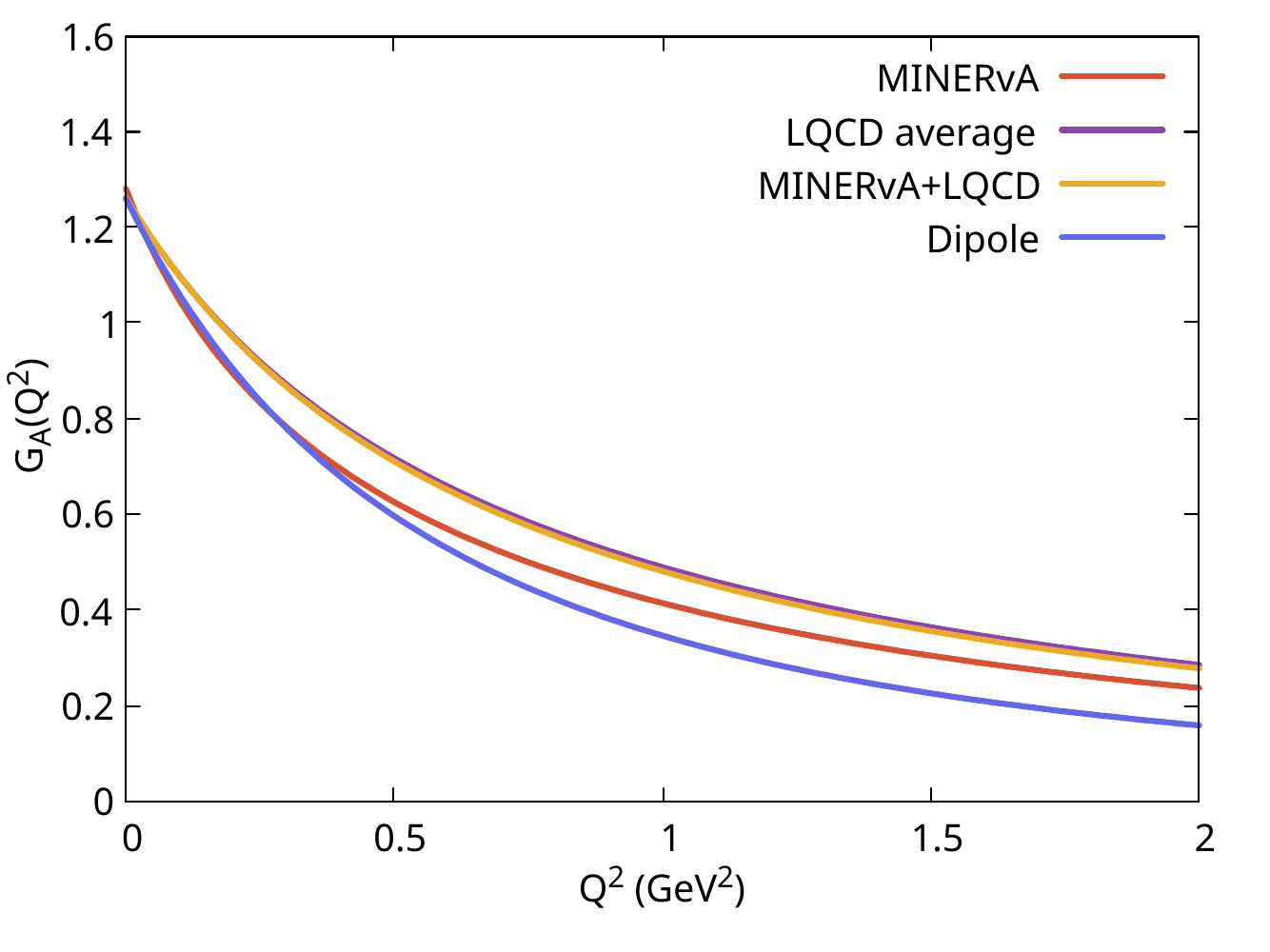}
\caption{Axial form factor $G_A(Q^2)$ as a function of $Q^2$ for the different parametrizations considered in this work: standard dipole with $M_A = 1.05$ GeV, MINERvA fit~\cite{MINERvA23}, and LQCD+MINERvA fit~\cite{Meyer25}. The LQCD average result from~\cite{Meyer26} is also shown for reference. All curves correspond to the central values of the parametrizations.}
\label{fig:ga}
\end{figure}

The one-body current operator is computed using the CC2 prescription~\cite{Udias93, Udias95, Martinez06}. In the case of neutrino-induced reactions, it contains both vector and axial components, the latter being proportional to the axial form factor $G_A(Q^{2})$.  The axial form factor has been commonly described using a standard dipole parametrization,
\be
G_A(Q^2)=\frac{g_A}{(1-Q^2/M_A^2)^2},
\ee
with $g_A=1.26$ and, here, we adopt an axial mass value of $M_A = 1.05$~GeV, extracted from neutrino-deuteron scattering and pion electroproduction data~\cite{Baker81, Miller82, Kitagaki83, Ahrens87, Bernard02}. However, recent LQCD calculations and measurements of antineutrino-proton scattering from the MINERvA collaboration suggest a significantly larger axial form factor, together with a shallower fall-off for $Q^2 > 0.3$~GeV$^2$~\cite{Bali20, Jang20, Park22, Meyer22, MINERvA23, Meyer25, Meyer26}. We investigate the impact of the axial form factor by presenting calculations obtained using the standard dipole parametrization, the MINERvA fit (for MINERvA data) given in~\cite{MINERvA23} and the combined LQCD+MINERvA fit reported by Meyer \textit{et al.}~\cite{Meyer25}.  These form factors are shown in Fig.~\ref{fig:ga}. For completeness, we also include the LQCD average result from~\cite{Meyer26}. All curves correspond to the central values of the respective parametrizations, which are the ones implemented in our model, while the associated uncertainties reported in the original works are not considered here.

\begin{figure}[hbtp]
\includegraphics[width=0.8\columnwidth,angle=0]{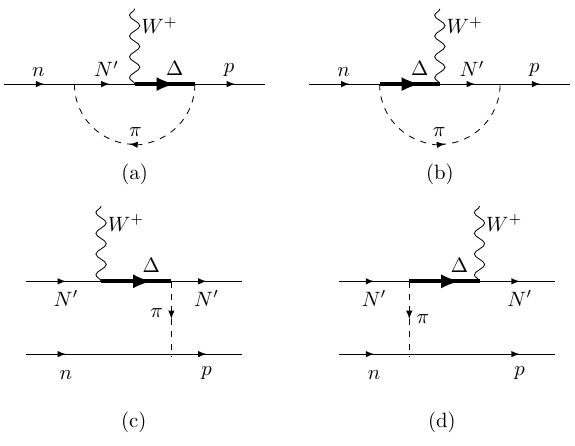}
\caption{$\Delta$-resonance diagrams contributing to two-body meson-exchange currents with a final one-particle-one-hole state. $N'$ denotes the intermediate bound-nucleon state, which can be either a neutron or a proton.}
\label{fig:delta_diagrams}
\end{figure}

\begin{figure}[hbtp]
\includegraphics[width=\columnwidth,angle=0]{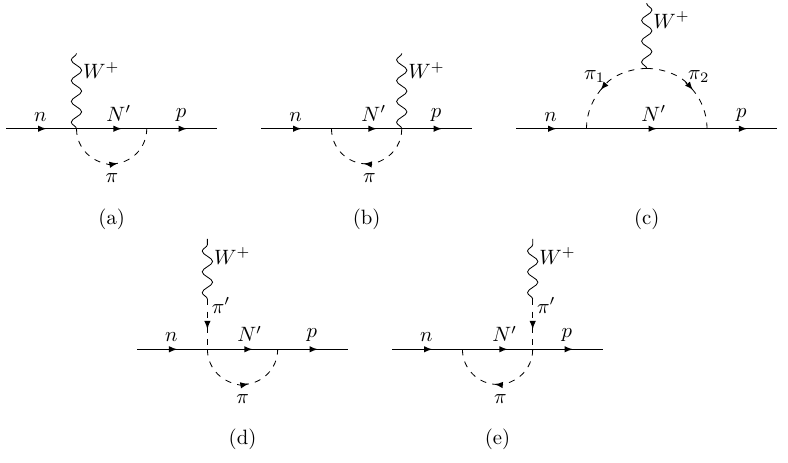}
\caption{ChPT background diagrams contributing to two-body meson exchange currents with a final one-particle one-hole state: contact term [(a) and (b)], pion-in-flight (c) and pion pole [(d) and (e)]. $N'$ denotes the intermediate bound-nucleon state, which can be either a neutron or a proton. }
\label{fig:chpt_diagrams}
\end{figure}

The two-body meson-exchange current contribution includes pion-exchange diagrams leading to a 1p1h final state. These comprise the $\Delta$-resonance mechanism, corresponding to the electroweak excitation of the $\Delta(1232)$ resonance followed by its decay into an $N\pi$ pair, with the pion being reabsorbed by another nucleon, as well as non-resonant contributions derived from a chiral perturbation theory (ChPT) Lagrangian of the pion-nucleon system~\cite{Scherer12}, shown in Figs.~\ref{fig:delta_diagrams} and \ref{fig:chpt_diagrams}, respectively. 1p1h excitations occur through two-body meson-exchange currents when one of the two nucleons involved remains bound to the nucleus. We refer to this nucleon as an intermediate bound-nucleon state. In this work, this intermediate nucleon, which is summed up to get the full 1p1h MEC contribution, is described as a free Dirac spinor in a relativistic Fermi gas with modified mass and energy, accounting for its interaction with a mean-field potential. This has been shown to provide a very good approximation to the full sum over all bound single-particle states obtained from the corresponding mean-field calculation~\cite{Franco-Munoz23,Franco-Munoz25}.  The attractive scalar potential then leads to an effective nucleon mass,
\be
M_N^*= \alpha M_N < M_N,
\ee
with $\alpha=0.8$.  Meanwhile, the vector potential produces a repulsive potential, $E_{v}$, which is added to the on-shell energy to obtain the modified nucleon energy,
\be
E^*=E + E_v,
\ee
where $E=\sqrt{p^2 + (M_N^*)^2}$ and $E_v=141$~MeV. The values of the effective mass and vector energy have been set following Refs.~\cite{Amaro18,Martinez-Consentino21,Ivanov24}. The explicit expressions of the operators and further details can be found in \cite{TaniaPhD}.

It is worth mentioning that, while all non-resonant contributions are derived from the ChPT Lagrangian~\cite{Scherer12}, this model does not provide the coupling with the $\Delta$-resonance, which has to be introduced \textit{ad hoc} from the formalism for the coupling to a spin-$3/2$ particle. Being separated pieces of the Lagrangian that are not derived from a common framework, the relative phase between the $\Delta$-resonant and ChPT terms is not unambiguously defined. In the case of single real-pion production, hints on the relative phase can be inferred assuming Watson's theorem~\cite{Watson52,Alvarez-Ruso16,Hooft26}, which follows from unitarity and time reversal invariance. In principle, one could consider employing this phase in the calculations involving two-body meson-exchange currents with a 1p1h final state.  However, the phase is a function of $W$. Therefore, for the processes we deal with in this work (virtual pions), this $\Delta$-resonance energy is significantly different from the values of $W$ relevant for real single pion production.  Most important, Watson’s theorem constrains the relative phase only up to an overall sign. That is, the phase can be either $\phi(W)$ or $\phi(W) + \pi$.  Thus, there is no theoretical indication on which to lean to determine the choice of a positive or negative sign in the $\Delta$-resonance amplitudes employed in this work for the two-body operators. The choice of this sign, however, has a direct impact on the interference with the one-body current contribution, thereby modifying the transverse response. The convention adopted in this work produces a constructive interference of the one-body and $\Delta$-resonance contributions, resulting in an enhancement of the transverse channel, whereas the opposite-sign choice would lead to a destructive interference between the one-body and $\Delta$-resonance terms, yielding a net decrease of the transverse response once the ChPT contribution is included. Our choice of sign is supported and motivated by the agreement with electromagnetic experimental data~\cite{Franco-Munoz23,Franco-Munoz25}. This is further justified by previous studies on the effect of two-body currents on the transverse response in inclusive scattering.  The enhancement of the transverse response has been observed within several frameworks, such as the relativistic Fermi gas model~\cite{Dekker94}, a non-relativistic mean-field-based model that incorporates final state interactions~\cite{VanderSluys95,Niewczas-PhD}, non-relativistic \textit{ab initio} Green’s Function Monte Carlo (GFMC) calculations~\cite{Carlson02,Lovato15,Lovato16,Lovato20,Andreoli22}, a variational Monte Carlo method with the short-time approximation~\cite{Andreoli24}, and the factorized spectral function formalism~\cite{Lovato23}, among others. Notably, GFMC calculations, have confirmed that the same choice of sign for the $\Delta$-resonance contributions that leads to an increase in the transverse response also yields improved predictions of magnetic moments and M1 transitions in $A \leq 9$ nuclei~\cite{Pastore13}. In these cases, two-body currents introduce significant corrections to the calculated observables, bringing them into excellent agreement with experimental data, further supporting our sign. Additionally, the enhancement in the transverse response persists in GFMC calculations even when employing simplified ground-state wave functions that neglect strong tensor correlations~\cite{Carlson02}, thus reinforcing the expected effect of this contribution in our approach, where correlations are not explicitly introduced via operators, but are effectively accounted for with the representation of the semi-phenomenological spectral function. This result contrasts, however, with earlier expectations based on Fermi-gas calculations of nuclear matter~\cite{Carlson94, Fabrocini97, Amaro02, Amaro03, Casale25}. In these works, an opposite sign for the $\Delta$-resonant contribution was adopted, resulting in a suppression of the transverse response. As a consequence, an alternative mechanism was required to account for the enhancement in the transverse channel shown by the experimental data. 
In particular, in~\cite{Fabrocini97}, it is attributed to the interplay between two-body currents and the presence of strong tensor correlations in the ground-state wave function.

\section{Results}\label{sec:results}

Our model is implemented in the NEUT~\cite{Hayato21} event generator as an extension of the framework described in Ref.~\cite{McKean25}, now including the contribution from two-body meson-exchange currents and allowing for different axial form factor parametrizations. The implementation relies on precomputed hadron tensors, with the one-body vector-vector (VV), vector-axial (VA), axial-axial (AA), and two-body components stored separately. Within this setup, the dependence on the axial form factor enters through
\begin{equation}
    \begin{split}
        & H_{1b,\text{VA}} \propto G_{A}(Q^{2}), \\
        & H_{1b,\text{AA}} \propto G_{A}^{2}(Q^{2}),
    \end{split}
\end{equation}
where the default choice corresponds to the dipole parametrization and it can be reweighted to a different form, $G_{A}^{'}$, as 
\begin{equation}
    \begin{split}
        & H_{1b,\text{VA}}' = \frac{G_{A}'(Q^{2})}{G_{A}(Q^{2})} H_{1b,\text{VA}}, \\
        & H_{1b,\text{AA}}' = \frac{G_{A}'^{2}(Q^{2})}{G_{A}^{2}(Q^{2})} H_{1b,\text{AA}}.
    \end{split}
\end{equation}
We note that the two-body component contains both the pure two-body contribution and the interference term between one- and two-body currents. Since the vector-axial separation is not performed in the two-body sector, modifying the axial form factor in the interference term is not feasible within the present implementation. However, implementing such a refinement would require storing additional hadron tensor components, increasing the memory footprint and the size of the tables used within the generator. Given that the overall two-body contribution enhances the cross section by up to about 20\%~\cite{Franco-Munoz23,Franco-Munoz25}, the limited impact expected from modifying the interference term does not justify the additional computational and storage cost. As a result, the two-body contribution is kept fixed for all axial form factor parametrizations and corresponds to that computed with the default dipole form.

Within this implementation, the total cross section as a function of neutrino energy is required in NEUT to normalize the event rate for a given flux, and therefore a dedicated cross section table must be produced for each hadron tensor input. In the present work, this also implies generating a separate table for each axial form factor parametrization considered. For completeness, the corresponding total cross sections are shown in Appendix \ref{app:tot_xsec}.

NEUT samples for the EDRMF potential were prepared with NEUT's intra-nuclear cascade turned on. The two-particle-two-hole (2p2h) channel is based on Refs.~\cite{Nieves11, Nieves12b} with corrections from Ref.~\cite{Bourguille:2021} and the pion production model is the Rein-Seghal model with added lepton mass corrections~\cite{Rein:1980wg, PhysRevD.76.113004, LepMassEff-Graczyk-Sobczyk-PhysRevD.77.053003}. The 2p2h channel also contains the three-particle-three-hole contribution as described in Ref.~\cite{NuWro_3p3h:PhysRevD.111.036032}. 
Also produced as a benchmark is the NEUT SF model based on Ref.~\cite{Benhar08} and the Nieves \textit{et al.} model based on Refs.~\cite{Nieves11,Nieves12b}, denoted as SF and N1p1h respectively. All samples are processed using the NUISANCE~\cite{Nuisance} framework. 
In order to decide whether a model prediction is in agreement with a measurement, we report the $\chi^{2}$ along with the associated $p$-values. The $\chi^{2}$ definition used in this work is given by
\begin{equation}
    \chi^{2} = \sum_{i,~j} \big( D - M \big)_{i} \big(\text{Cov}^{-1} \big)_{ij} \big( D - M\big)_{j}.
\end{equation}
Here, $i$ and $j$ indicate a particular bin of a measurement. $D$ and $M$ represent the data and MC value in a given bin respectively. The inverse covariance matrix is given by $\text{Cov}^{-1}$ and has dimensions $i \times j$. The associated $p$-value is calculated by 
\begin{equation}
    p = 1 - F_{\chi^{2}}(\chi^{2}, N_{\text{d.o.f}}),
\end{equation}
where $F$ is the cumulative distribution function for a given $\chi^{2}$ value and number of degrees of freedom $ N_{\text{d.o.f}}$.  The $p$-value is bounded in the interval $[0,1]$, with $p \approx 0$ indicating that the model is rejected by the $\chi^{2}$ test. In this work, models with $p \geq 0.05$ are considered to be compatible with the data.

\subsection{Comparison to T2K inclusive and semi-inclusive measurements}

\begin{figure*}[t]
    \centering

    \begin{minipage}[b]{0.45\textwidth}
        \centering
        \includegraphics[width=\textwidth]{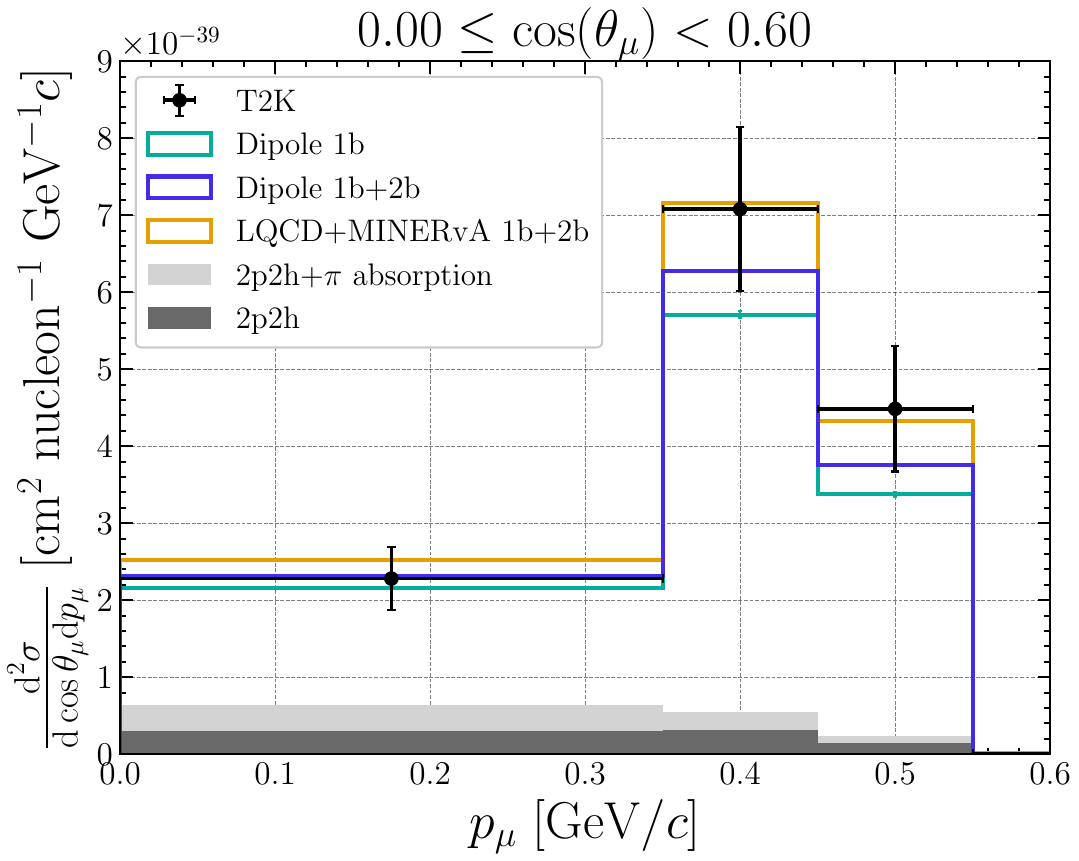}
        \label{fig:T2K:incl:a}
    \end{minipage}
    \hfill
    \begin{minipage}[b]{0.45\textwidth}
        \centering
        \includegraphics[width=\textwidth]{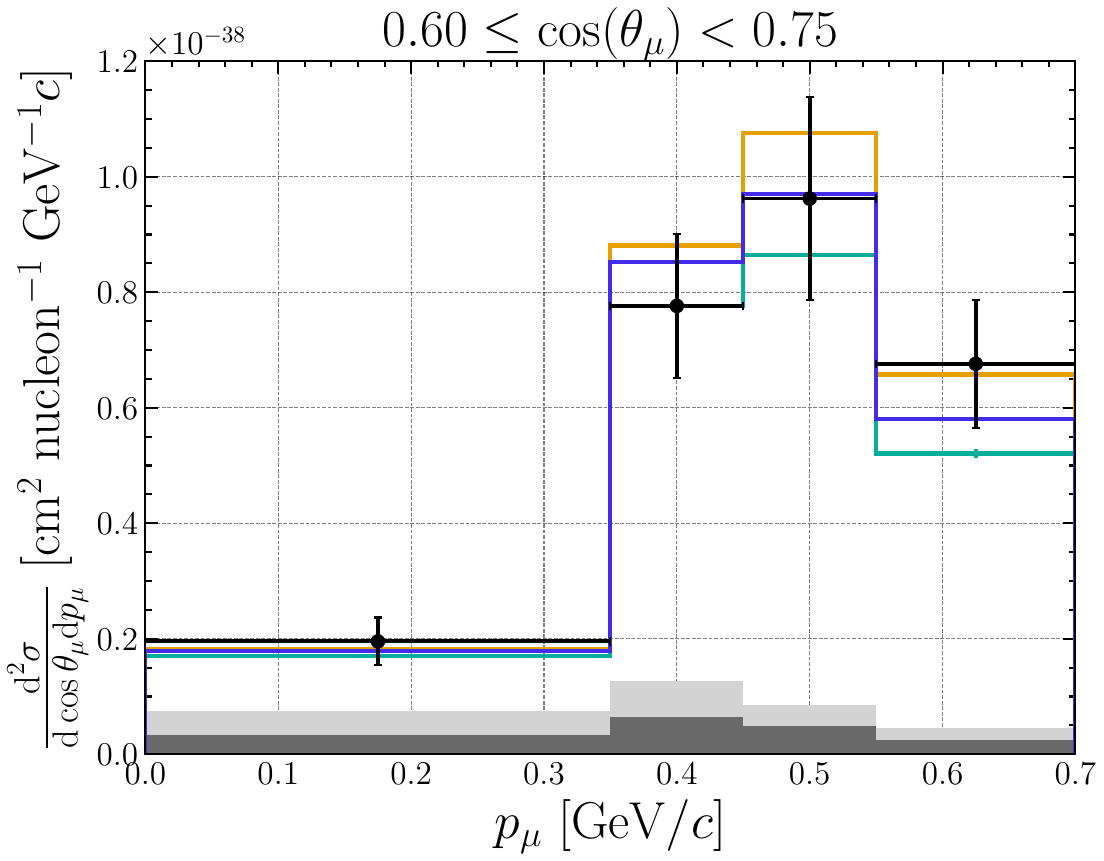}
        \label{fig:T2K:incl:b}
    \end{minipage}

    \vspace{0.3cm}

    \begin{minipage}[b]{0.45\textwidth}
        \centering
        \includegraphics[width=\textwidth]{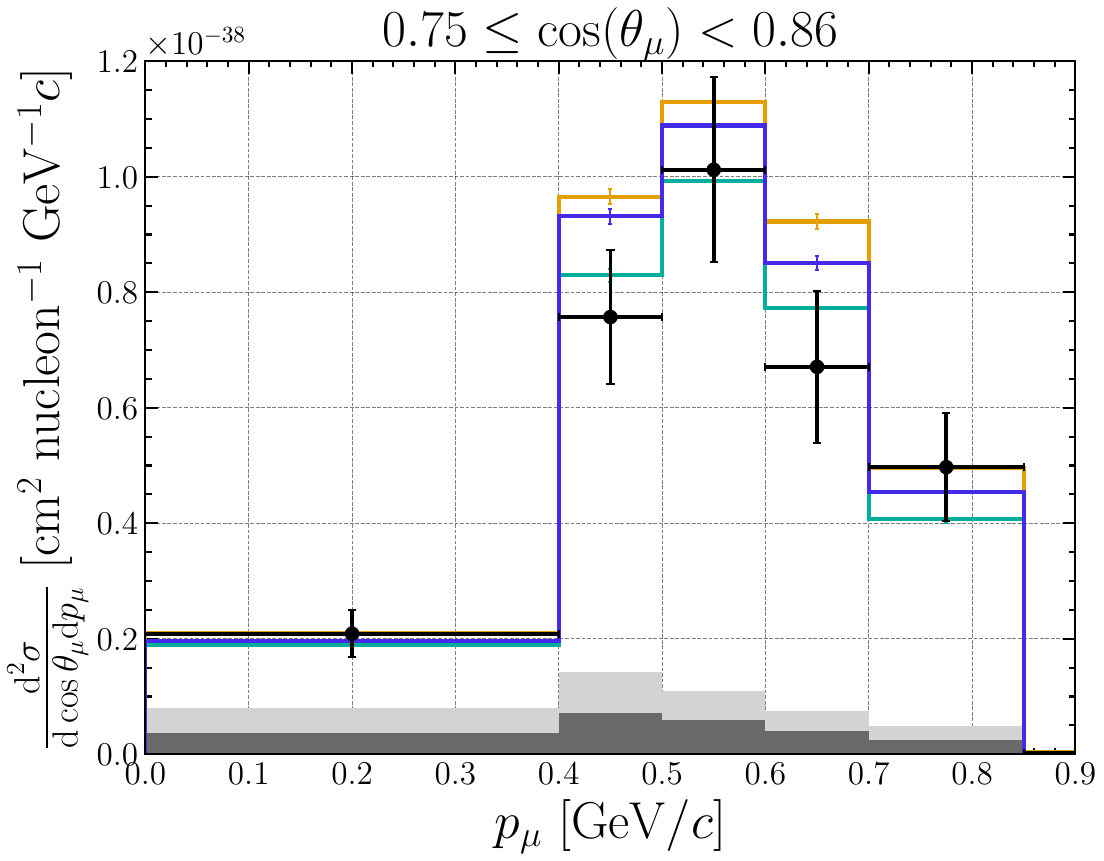}
        \label{fig:T2K:incl:c}
    \end{minipage}
    \hfill
    \begin{minipage}[b]{0.45\textwidth}
        \centering
        \includegraphics[width=\textwidth]{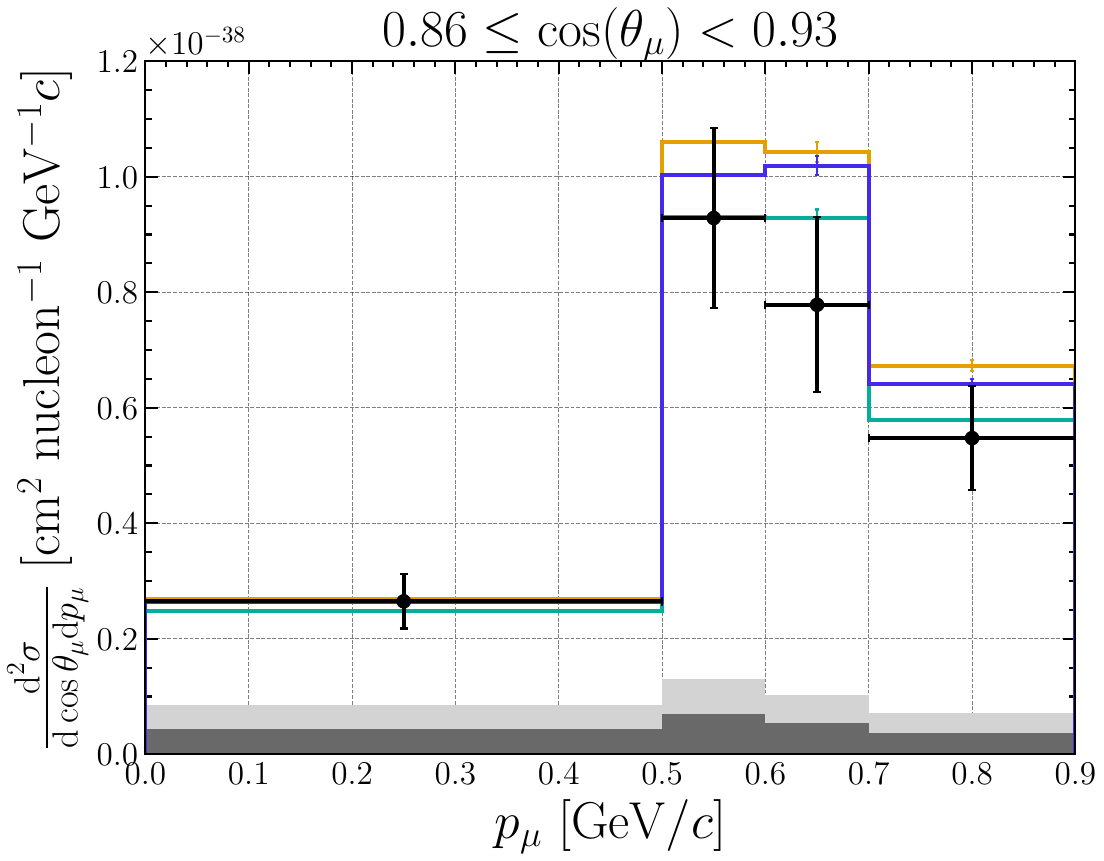}
        \label{fig:T2K:incl:d}
    \end{minipage}

    \vspace{0.3cm}

    \begin{minipage}[b]{0.45\textwidth}
        \centering
        \includegraphics[width=\textwidth]{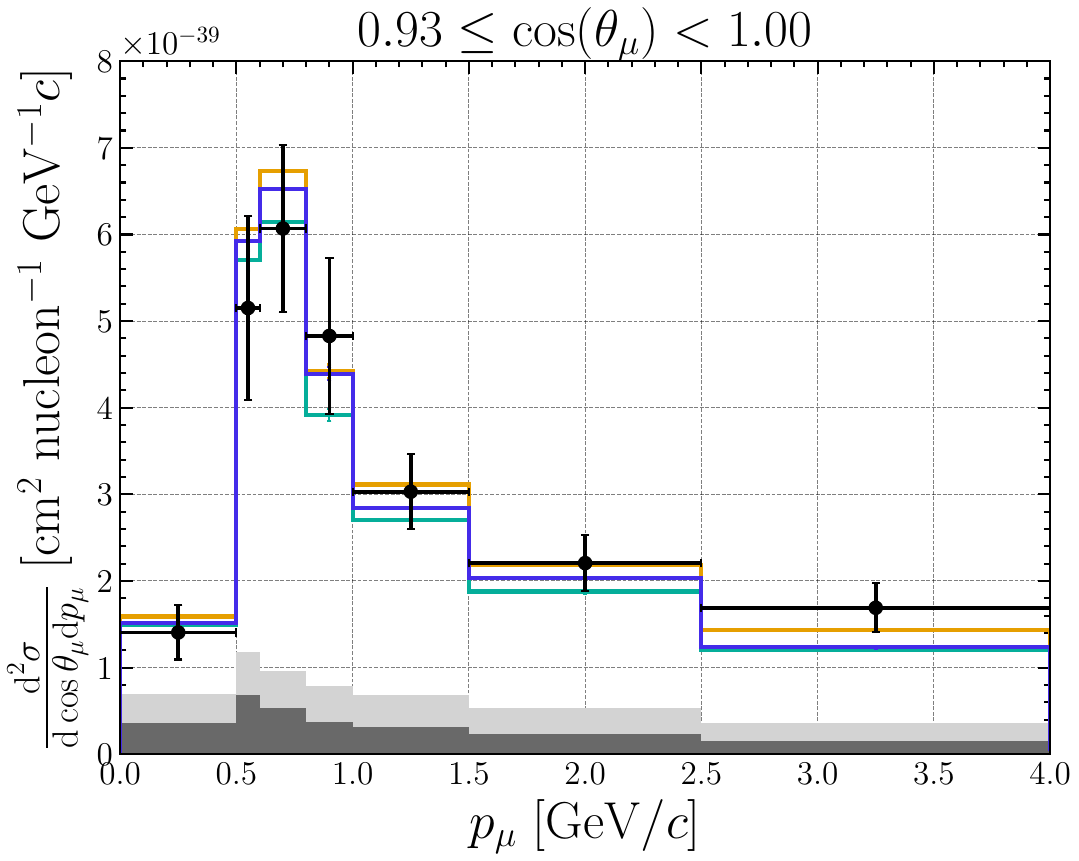}
        \label{fig:T2K:incl:e}
    \end{minipage}

    \caption{Comparison between the NEUT EDRMF model using the dipole axial form factor with one-body (green) and one- and two-body (blue) current contributions, and the LQCD+MINERvA axial form factor with one- and two-body current contributions (orange). The inclusive measurement is taken from Ref.~\cite{T2K20b}. The 2p2h contribution is shown in light grey, including pion absorption, and in dark grey, where pion absorption is not included.}
    \label{fig:T2K:incl}
\end{figure*}

In Fig~\ref{fig:T2K:incl}, we present our results for the CC$0\pi$ double-differential inclusive cross-section measurement of Ref.~\cite{T2K20b}, considering both the dipole and the LQCD+MINERvA parametrizations of the axial form factor. CC0$\pi$ scattering refers to CC interactions in which no pions are detected in the final state, while the number of outgoing protons can be either restricted or not restricted. For the data shown in Fig.~\ref{fig:T2K:incl}, the number of protons is unrestricted. Our simulation includes CCQE interactions, multi-nucleon excitations and pion production events in which the pion is reabsorbed in the nuclear medium~\cite{T2K16}. Differential cross sections are plotted against the muon momentum, $p_{\mu}$, and each panel corresponds to a different bin in the muon scattering angle, $\cos\theta_{\mu}$. 

The recent axial form factor extracted from the combined LQCD+MINERvA analysis from \cite{Meyer25,Meyer26} leads to a systematic enhancement of the cross section with respect to the standard dipole parametrization. This effect is more pronounced at larger
muon scattering angles, where differences in the momentum-transfer dependence of the two parametrizations become more relevant. As the scattering becomes more forward, the difference between the two form factor descriptions gradually diminishes. Two-body meson-exchange currents also produce an overall enhancement of the cross section.  This effect can be understood from the structure of charged-current cross sections, which can be decomposed into charge-charge, charge-longitudinal, longitudinal-longitudinal, and two transverse ($T$, $T'$) hadronic responses~\cite{Amaro05a}. Two-body currents enhance the magnitude of both $T$ and $T'$ responses, while leaving the charge and longitudinal components essentially unchanged. The transverse $T$ channel dominates over the full kinematic range in charged-current neutrino scattering, therefore, the contribution of two-body currents translates into a sizeable increase of the total cross section. Once averaged over the neutrino flux, this contribution remains approximately constant across the different $\cos{\theta_\mu}$ bins, yielding an increase of up to about 12\%. It is also worth noting that the main difference between charged-current neutrino and antineutrino cross sections appears in the leptonic tensor and affects the sign of the $T'$ contribution, which adds up for neutrino scattering but reduces strength in the antineutrino case. Therefore, the relative increase in the cross sections due to two-body currents is larger for neutrinos compared to antineutrinos as, for the latter, the enhancement of the $T$ and $T'$ responses has opposite signs, diminishing the overall effect.

Table~\ref{tab:T2K:incl_chi2} reports the $\chi^{2}$ values and corresponding $p$-values obtained for the different axial form factor parametrizations and for calculations performed with and without two-body meson-exchange currents. For comparison, results from the NEUT SF and N1p1h CCQE models are also included.  Overall, the measurement is well reproduced by our framework across all considered scenarios.  Although the inclusion of two-body currents and the use of the LQCD+MINERvA axial form factor lead to slightly larger $\chi^{2}$ values in comparison to the one-body dipole case, all configurations remain statistically acceptable according to their $p$-values. The NEUT SF and N1p1h models are likewise accepted, with the N1p1h configuration exhibiting near-perfect agreement.  However, both approaches rely on simplified descriptions of nuclear dynamics. The N1p1h model is based on the relativistic Fermi gas, which oversimplifies the nuclear structure. Meanwhile, spectral function approaches provide a more realistic description of the nuclear ground state, but cross sections are typically computed within a factorized formalism, which is not always a reliable approximation~\cite{Caballero98a,Nikolakopoulos19} and prevents the consistent inclusion of final-state interactions~\cite{Gonzalez-Jimenez19}.

\begin{table} [htbp]
  \caption{$\chi^{2}/N_{\text{d.o.f}}$ and $p$-values for the T2K inclusive measurement from Ref.~\cite{T2K20b}. Bold rows indicate configurations that are accepted based on the assertion that $p$-value $\ge 0.05$.}
  \label{tab:T2K:incl_chi2}
  \centering
  \begin{tabular*}{1.0\columnwidth}{@{\extracolsep{\fill}}c c c c }
    \hline \hline
    Sample & Model & $\chi^{2}/N_{\text{d.o.f}}$ & $p$-value \\
    \hline
    $\frac{\text{d}^{2}\sigma}{\text{d}\cos \theta_{\mu}\text{d}p_{\mu}}$ &\textbf{Dipole 1b} & $\mathbf{29.82/29}$ & $\mathbf{0.42}$ \\
    $\frac{\text{d}^{2}\sigma}{\text{d}\cos \theta_{\mu}\text{d}p_{\mu}}$ &\textbf{Dipole 1b+2b} & $\mathbf{39.44/29}$ & $\mathbf{0.09}$ \\
    \makecell{$\frac{\text{d}^{2}\sigma}{\text{d}\cos \theta_{\mu}\text{d}p_{\mu}}$} & \makecell{\textbf{LQCD+Minerva}\\\textbf{ 1b+2b}} & \makecell{$\mathbf{32.89/29}$} & \makecell{$\mathbf{0.28}$} \\
    $\frac{\text{d}^{2}\sigma}{\text{d}\cos \theta_{\mu}\text{d}p_{\mu}}$ & \textbf{SF} & $\mathbf{32.43/29}$ & $\mathbf{0.30}$ \\
    $\frac{\text{d}^{2}\sigma}{\text{d}\cos \theta_{\mu}\text{d}p_{\mu}}$ & \textbf{N1p1h} & $\mathbf{12.83/29}$ & $\mathbf{1.00}$ \\
    \hline \hline
  \end{tabular*}

\end{table}

\begin{figure*}[t]
    \centering
    
    \begin{minipage}[b]{0.45\textwidth}
        \centering
        \includegraphics[width=\textwidth]{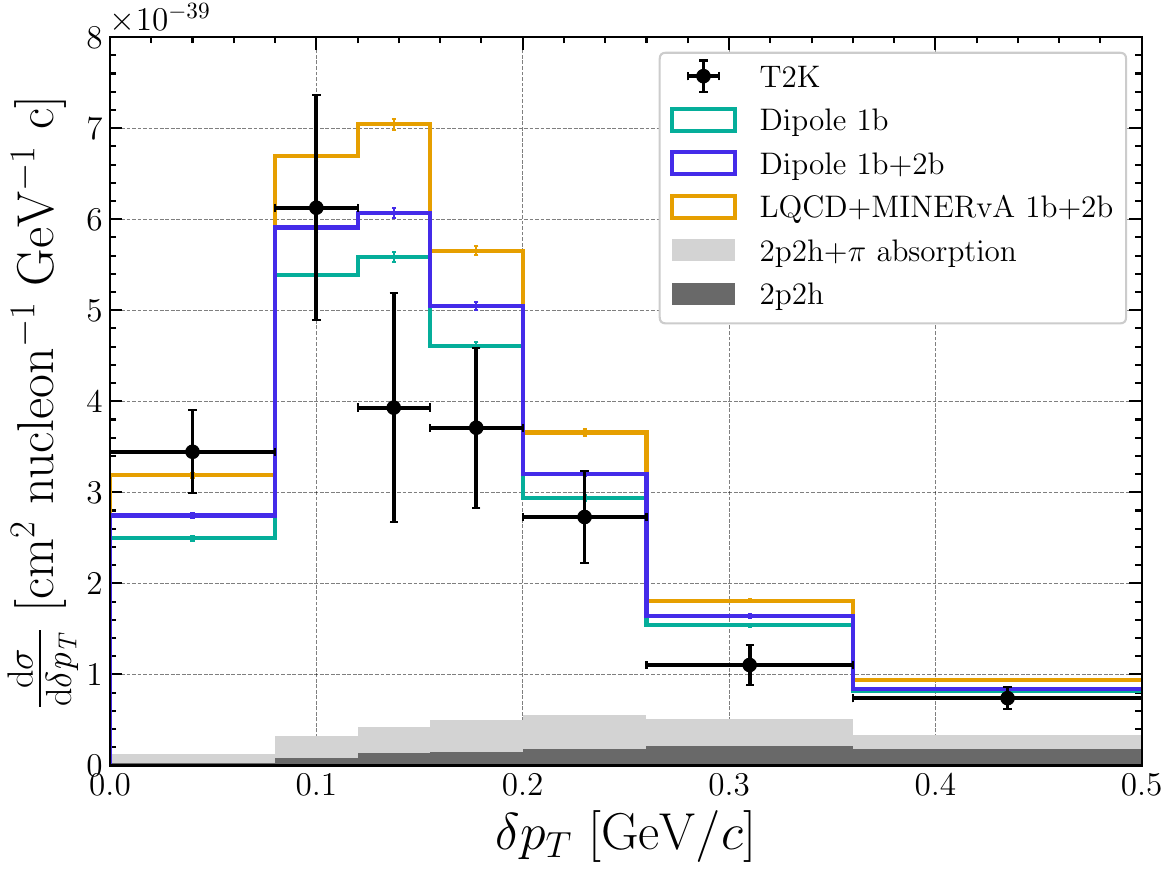}
    \end{minipage}
    \hfill
    \begin{minipage}[b]{0.45\textwidth}
        \centering
        \includegraphics[width=\textwidth]{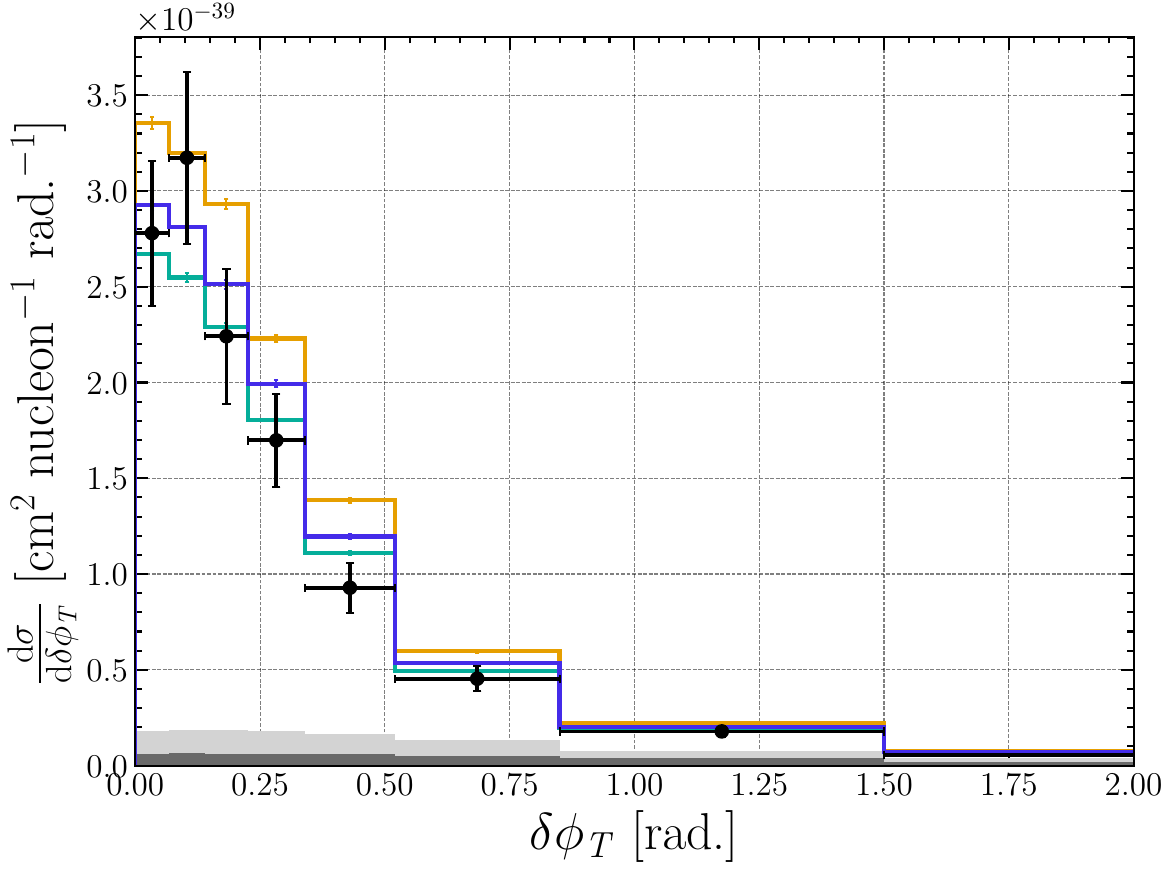}
    \end{minipage}
    
    \vspace{0.3cm}
    
    \begin{minipage}[b]{0.45\textwidth}
        \centering
        \includegraphics[width=\textwidth]{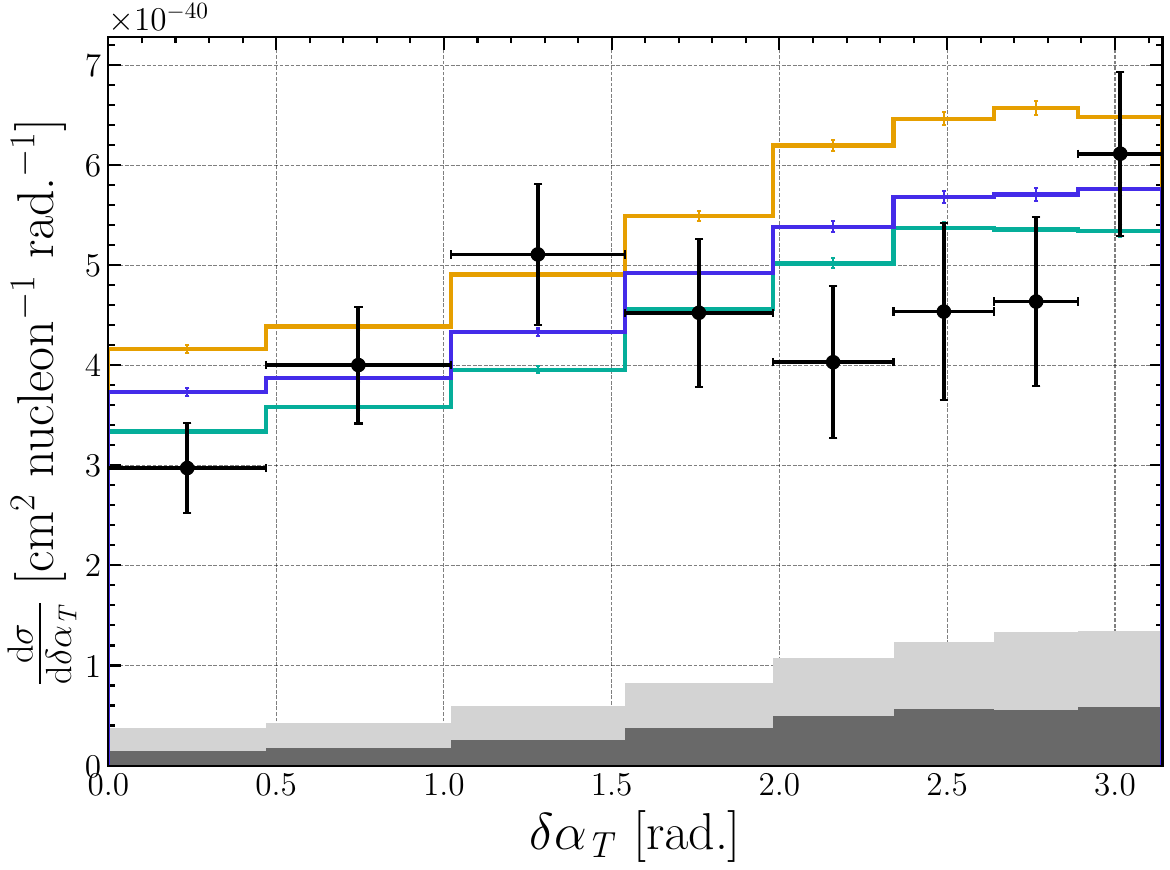}
    \end{minipage}
    
    \caption{Same as Fig.~\ref{fig:T2K:incl}, but comparing against transverse kinematic imbalance measurements from Ref.~\cite{T2K18}.}
    \label{fig:T2K:TKI}
\end{figure*}

In Fig.~\ref{fig:T2K:TKI}, we present our results for the T2K CC$0\pi$Np measurement~\cite{T2K18}, expressed in terms of transverse kinematic imbalance (TKI) variables described in Ref.~\cite{Pickering:2016}.  Table~\ref{tab:T2K:TKI_chi2} shows the corresponding $\chi^{2}$ and $p$-values, together with the results from the SF and N1p1h CCQE models. For the $\delta p_T$ variable, our one-body calculation with the dipole axial form factor yields the best $\chi^{2}$ among our configurations, although it remains larger than that obtained with the alternative NEUT models. The discrepancy in both dipole cases can be traced back to an underprediction of the first bin at low $\delta p_T$. In this region, where CCQE dominates, the  LQCD+MINERvA axial form factor configuration provides an improved description. However, this improvement at low $\delta p_T$ is accompanied by a tendency to overestimate the cross section at larger values of $\delta p_T$. A similar pattern is observed for the $\delta\phi_T$ variable. Our model generally exhibits larger $\chi^{2}$ values, although in this case the agreement improves when two-body currents are included together with the dipole axial form factor. The use of the LQCD+MINERvA parametrization leads instead to a systematic overestimation across several bins, resulting in a poorer overall description. For $\delta \alpha_T$, the situation differs. Here, our one-body dipole calculation provides the best agreement, followed closely by the NEUT SF model and our dipole calculation that includes two-body currents, which yield comparable $\chi^{2}$ values. The N1p1h model and the LQCD+MINERvA axial form factor calculation show a comparatively poorer agreement. However, it is important to note that none of the considered models fully reproduces the slightly peaked structure observed in the data between 1 and 2 radians.

\begin{table} [htbp]
  \caption{$\chi^{2}/N_{\text{d.o.f}}$ and $p$-values for the T2K TKI measurement. Bold rows indicate configurations that are accepted based on the assertion that $p$-value $ \ge 0.05$.}
  \label{tab:T2K:TKI_chi2}
  \centering
  \begin{tabular}{l c c c}
    \hline
    \hline
    Sample & Model & $\chi^{2}/N_{\text{d.o.f}}$ & $p$-value \\
    \hline
    $\delta p_{T}$& Dipole 1b & 33.40/8 & 0.00 \\
    $\delta p_{T}$& Dipole 1b+2b & 37.14/8 & 0.00 \\
    $\delta p_{T}$& LQCD+Minerva 1b+2b & 48.93/8 & 0.00 \\
    $\mathbf{\delta p_{T}}$& \textbf{SF} & $\mathbf{12.24/8}$ & $\mathbf{0.14}$ \\
    $\mathbf{\delta p_{T}}$& \textbf{N1p1h} & $\mathbf{8.74/8}$ & $\mathbf{0.36}$ \\
    \hline
    $\delta \phi_{T}$& Dipole 1b & 18.12/8 & 0.02 \\
    $\delta \phi_{T}$& Dipole 1b+2b & 17.11/8 & 0.03 \\
    $\delta \phi_{T}$& LQCD+Minerva 1b+2b & 25.39/8 & 0.00 \\
    $\mathbf{\delta \phi_{T}}$& \textbf{SF} & $\mathbf{7.19/8}$ & $\mathbf{0.52}$ \\
    $\mathbf{\delta \phi_{T}}$& \textbf{N1p1h} & $\mathbf{11.83/8}$ & $\mathbf{0.16}$ \\
    \hline
    $\delta \alpha_{T}$ & Dipole 1b & 18.20/8 & 0.02 \\
    $\delta \alpha_{T}$ & Dipole 1b+2b & 20.19/8 & 0.01 \\
    $\delta \alpha_{T}$ & LQCD+Minerva 1b+2b & 31.83/8 & 0.00 \\
    $\delta \alpha_{T}$ & SF & 19.20/8 & 0.01 \\
    $\delta \alpha_{T}$ & N1p1h & 29.11/8 & 0.00 \\
    \hline
    \hline
  \end{tabular}
\end{table}

\subsection{Comparison to MINERvA semi-inclusive measurements}

\begin{figure*}[t]
    \centering
    \begin{minipage}[b]{0.45\textwidth}
        \centering
        \includegraphics[width=\textwidth]{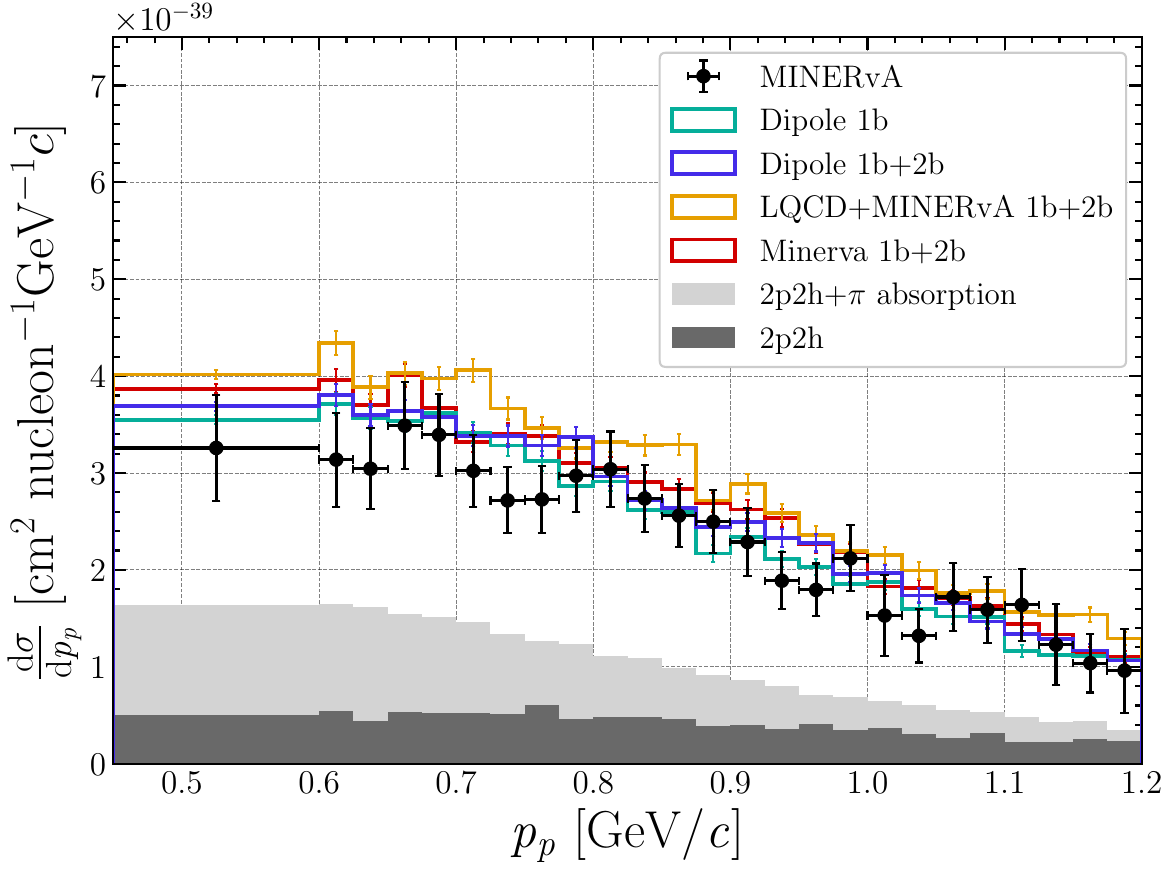}
    \end{minipage}
    \hfill
    \begin{minipage}[b]{0.45\textwidth}
        \centering
        \includegraphics[width=\textwidth]{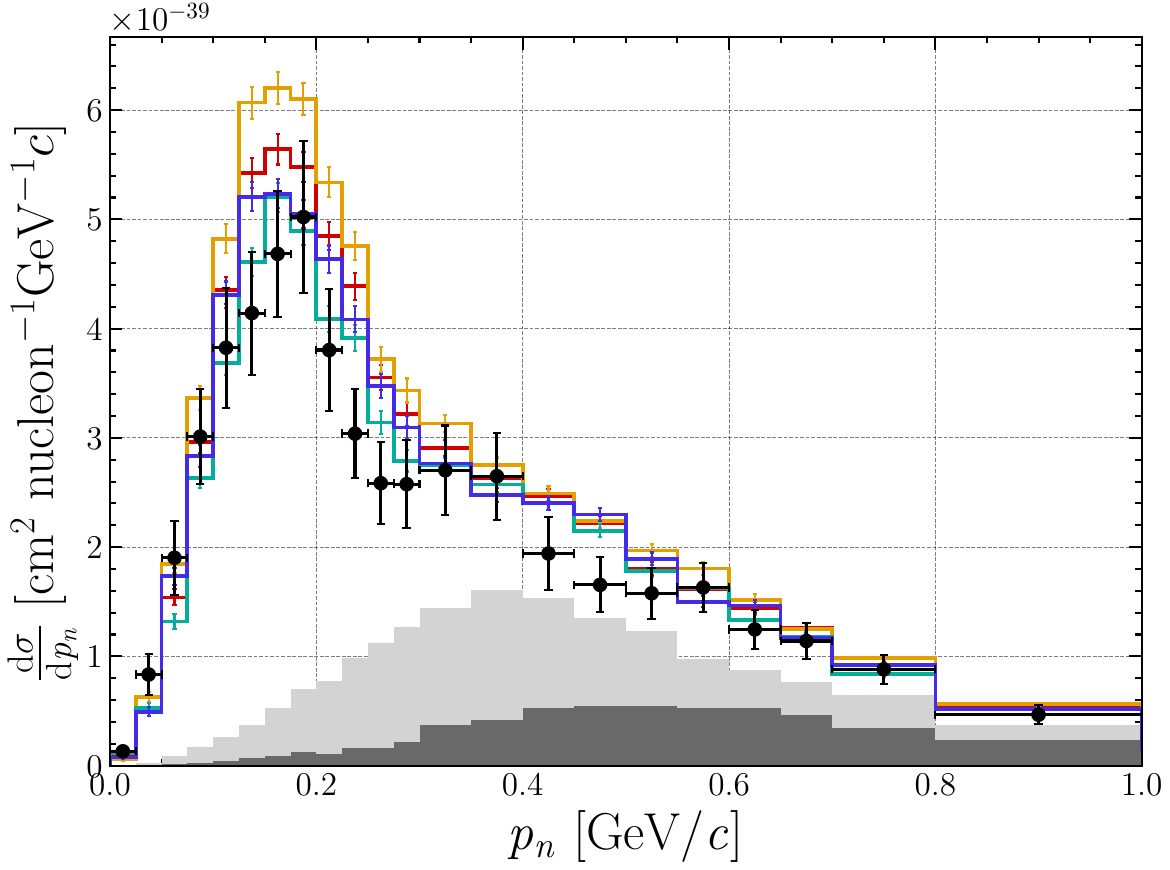}
    \end{minipage}
    
    \vspace{0.3cm}
    
    \begin{minipage}[b]{0.45\textwidth}
        \centering
        \includegraphics[width=\textwidth]{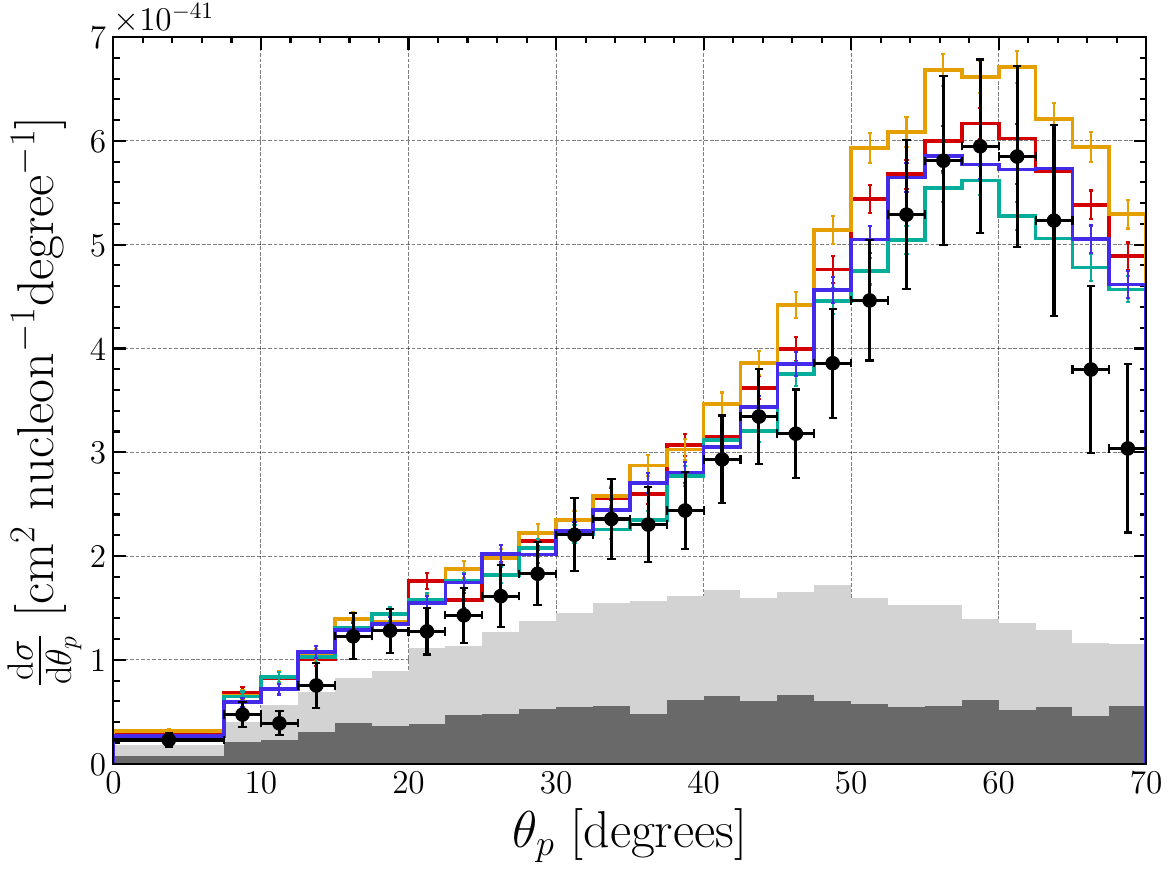}
    \end{minipage}
    \hfill
    \begin{minipage}[b]{0.45\textwidth}
        \centering
        \includegraphics[width=\textwidth]{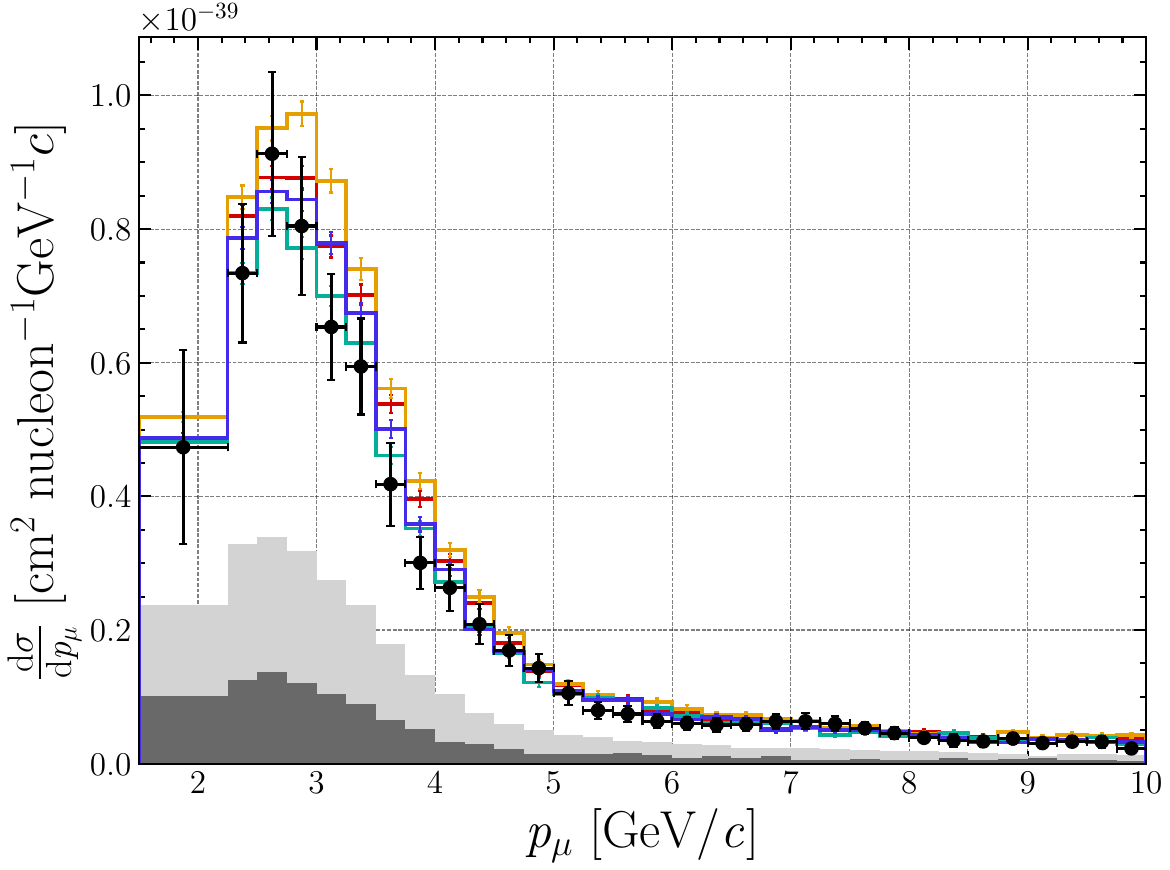}
    \end{minipage}
    
    \caption{Same as Fig.~\ref{fig:T2K:incl}, but comparing against MINERvA kinematic measurements from Refs.~\cite{Lu18, MINERvA20}. Included in red is also the result using the MINERvA axial form factor fit with one- and two-body current contributions.}
    \label{fig:Minerva:Kinematics}
\end{figure*}

The CC0$\pi$Np measurement from Ref.~\cite{Lu18} with a correction applied in Ref.~\cite{MINERvA20} is used to further benchmark the model. Fig.~\ref{fig:Minerva:Kinematics} shows the comparison against single-differential cross section measurements of kinematic variables for the muon, final-state proton and reconstructed initial-state neutron. In this case, in addition to the dipole and combined LQCD+MINERvA parametrizations considered previously, we also include the axial form factor obtained by the MINERvA collaboration from a fit to its data in~\cite{MINERvA23}. We note that the combined LQCD+MINERvA fit is largely dominated by the LQCD input (see Fig.~\ref{fig:ga}), as the latter provides a significantly tighter constraint on the uncertainty of the axial form factor. Hence, when comparing to MINERvA data, it is particularly relevant to explicitly include the MINERvA-only parametrization. Table~\ref{tab:Minerva_kinematic_chi2} shows the $\chi^{2}$ and $p$-values, together with the results from the SF and N1p1h CCQE models. For the proton momentum, $p_p$, the MINERvA axial form factor calculation shows the best agreement, followed by the dipole form factor case, both including two-body currents. The systematic enhancement shown in the comparison with T2K data by the LQCD+MINERvA calculation including two-body current contributions is also observed here.  For the initial-state neutron momentum, $p_n$, the difference in axial form factor and inclusion of two-body currents cause a drastic difference to the height of the distribution, 
which is strongly overestimated by the LQCD+MINERvA configuration. In contrast, the one-body dipole configuration predicts the peak well. The second peak-like structure observed around 0.3~GeV is captured well by all configurations but all models tend to overestimate the contribution between these two peaks. This is reflected in Table~\ref{tab:Minerva_kinematic_chi2}, in which all models are rejected.  For the scattered proton angle, $\theta_p$, all models overestimate the measurement at large angles, while the peak of the distribution around 60$^{\circ}$ is well predicted by all configurations. Interestingly, there exists a data point at small scattering angle that is already overestimated by the 2p2h with pion absorption contribution only. Removal of this point improves the $\chi^{2}$ values of all models and leads to the one-body dipole configuration being accepted.  The scattered muon momentum measurement further highlights the systematic enhancement from the LQCD+MINERvA form factor calculation with two-body currents. Despite this, the $\chi^{2}$ value for the LQCD+MINERvA configuration is still accepted. The one-body dipole configuration reports the best agreement followed by the NEUT SF. 

\begin{table} [htbp]
  \caption{$\chi^{2}/N_{\text{d.o.f}}$ and $p$-values for the MINERvA kinematic measurement. Bold rows indicate configurations that are accepted based on the assertion that $p$-value $\ge0.05$.}
  \label{tab:Minerva_kinematic_chi2}
  \centering
  \begin{tabular}{l c c c}
    \hline
    \hline
    Sample & Model & $\chi^{2}/N_{\text{d.o.f}}$ & $p$-value \\
    \hline
    $p_{p}$&Dipole 1b & 41.19/25 & 0.02 \\
    $\mathbf{p_{p}}$&\textbf{Dipole 1b+2b} & $\mathbf{34.35/25}$ & $\mathbf{0.10}$ \\
    $p_{p}$&LQCD+MINERvA 1b+2b & 63.29/25 & 0.00 \\
    $\mathbf{p_{p}}$&\textbf{MINERvA 1b+2b} & $\mathbf{25.81/25}$ & $\mathbf{0.42}$ \\
    ${p_{p}}$& SF & 73.33/25 & 0.00 \\
    ${p_{p}}$& N1p1h & 38.43/25 & 0.04 \\
    \hline
    $p_{n}^{\text{reco}}$&Dipole 1b & 46.96/24 & 0.00 \\
    $p_{n}^{\text{reco}}$&Dipole 1b+2b & 78.94/24 & 0.00 \\
    $p_{n}^{\text{reco}}$&LQCD+MINERvA 1b+2b & 71.72/24 & 0.00 \\
    $p_{n}^{\text{reco}}$& MINERvA 1b+2b & 59.17/24 & 0.00 \\
    $p_{n}^{\text{reco}}$& SF & 72.71/24 & 0.00 \\
    $p_{n}^{\text{reco}}$& N1p1h & 158.68/24 & 0.00 \\
    \hline
    $\theta_{p}$&Dipole 1b & 53.87/26 & 0.00 \\
    $\mathbf{\theta_{p}}$&\textbf{Dipole 1b+2b} & $\mathbf{34.88/26}$ & $\mathbf{0.11}$ \\
    $\theta_{p}$&LQCD+MINERvA 1b+2b & 71.22/26 & 0.00 \\
    $\theta_{p}$&MINERvA 1b+2b & 74.59/26 & 0.00 \\
    $\theta_{p}$& SF & 56.26/26 & 0.00 \\
    $\theta_{p}$& N1p1h & 52.77/26 & 0.00 \\
    \hline
    $\mathbf{p_{\mu}}$&\textbf{Dipole 1b} & $\mathbf{35.81/32}$ & $\mathbf{0.29}$ \\
    $p_{\mu}$&Dipole 1b+2b & 48.35/32 & 0.03 \\
    $\mathbf{p_{\mu}}$ & \makecell{\textbf{LQCD+Minerva}\\\textbf{ 1b+2b}} & \makecell{$\mathbf{45.99/32}$} & \makecell{$\mathbf{0.05}$} \\
    $\mathbf{p_{\mu}}$&\textbf{MINERvA 1b+2b} & \textbf{46.60/32} & \textbf{0.05} \\
    $\mathbf{p_{\mu}}$& \textbf{SF} & $\mathbf{37.76/32}$ & $\mathbf{0.22}$ \\
    $p_{\mu}$& N1p1h & 69.44/32 & 0.00 \\
    \hline
    \hline
  \end{tabular}
\end{table}

\begin{figure*}[t]
    \centering

    \begin{minipage}[b]{0.45\textwidth}
        \centering
        \includegraphics[width=\textwidth]{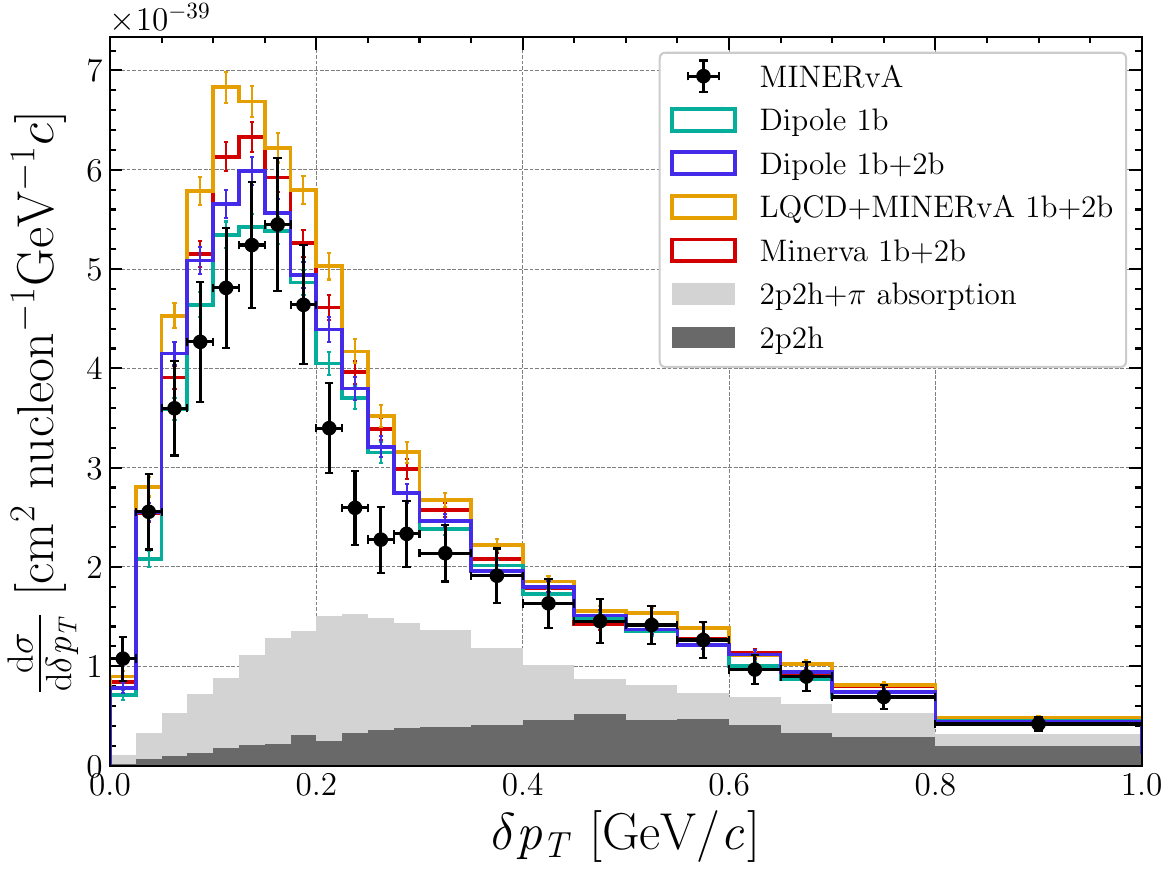}
    \end{minipage}
    \hfill
    \begin{minipage}[b]{0.45\textwidth}
        \centering
        \includegraphics[width=\textwidth]{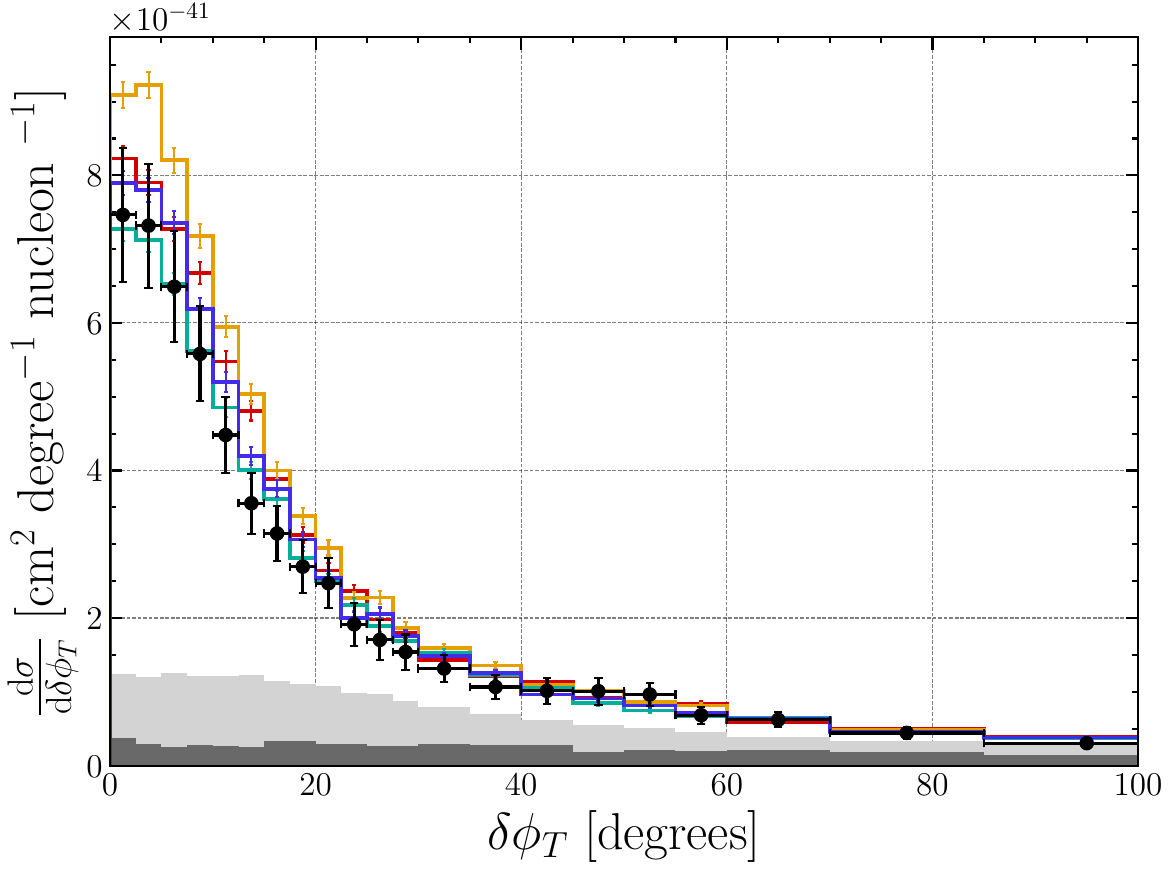}
    \end{minipage}

    \vspace{0.3cm}

    \begin{minipage}[b]{0.45\textwidth}
        \centering
        \includegraphics[width=\textwidth]{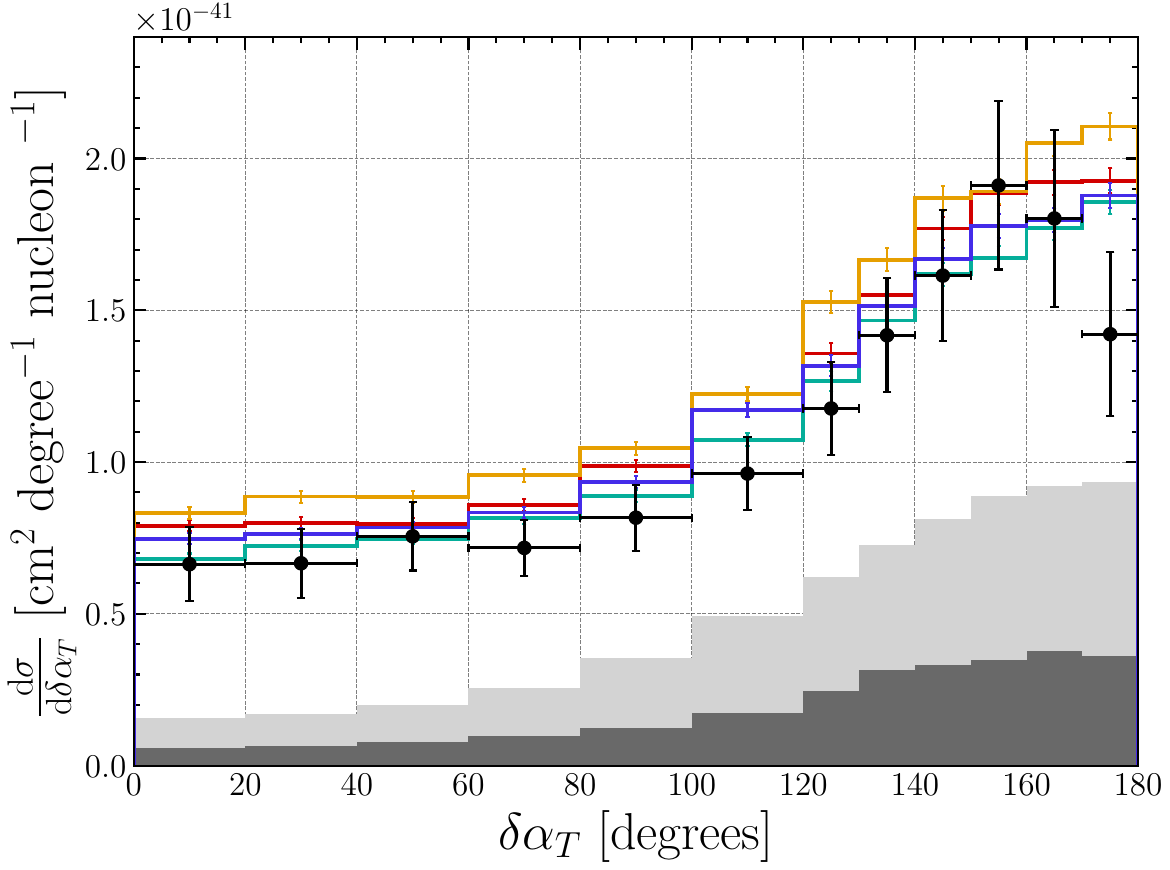}
    \end{minipage}

    \caption{Same as Fig.~\ref{fig:Minerva:Kinematics}, but comparing against MINERvA transverse kinematic imbalance measurements from Refs.~\cite{Lu18, MINERvA20}.}
    \label{fig:Minerva:TKI}
\end{figure*}

Fig.~\ref{fig:Minerva:TKI} shows the comparison against TKI variables. 
For $\delta p_T$, the peak is overestimated by all configurations that include the two-body currents and is only predicted well by the one-body dipole configuration. Between 0.2 and 0.3~GeV, all models overestimate the measurement. Meanwhile, at larger values of $\delta p_T$, all models are consistent with the measurement. As a result of this overestimation in many areas, all models are rejected based on the $\chi^{2}$ and $p$-values shown in Table~\ref{tab:Minerva_TKI_chi2}. $\delta \phi_T$ shows a similar trend in which the LQCD+MINERvA form factor result with two-body currents overestimates the measurement. All models are also rejected for this variable. Finally, for $\delta \alpha_T$, only the NEUT SF model is accepted. The last bin is overestimated by all configurations and removal of this bin results in the acceptance of the dipole configurations.

\begin{table} [htbp]
  \caption{$\chi^{2}/N_{\text{d.o.f}}$ and $p$-values for the MINERvA TKI measurement. Bold rows indicate configurations that are accepted based on the assertion that $p$-value $\ge0.05$.}
  \label{tab:Minerva_TKI_chi2}
  \centering
  \begin{tabular}{l c c c}
    \hline
    \hline
    Sample & Model & $\chi^{2}/N_{\text{d.o.f}}$ & $p$-value \\
    \hline
    $\delta p_{T}$&Dipole 1b & 51.03/24 & 0.00 \\
    $\delta p_{T}$&Dipole 1b+2b & 96.06/24 & 0.00 \\
    $\delta p_{T}$&LQCD+MINERvA 1b+2b & 100.85/24 & 0.00 \\
    $\delta p_{T}$& MINERvA 1b+2b & 90.48/24 & 0.00 \\
    $\delta p_{T}$& SF & 94.86/24 & 0.00 \\
    $\delta p_{T}$& N1p1h & 127.38/24 & 0.00 \\
    \hline
    $\delta \phi_{T}$&Dipole 1b & 45.34/23 & 0.00 \\
    $\delta \phi_{T}$&Dipole 1b+2b & 47.13/23 & 0.00 \\
    $\delta \phi_{T}$&LQCD+MINERvA 1b+2b & 67.49/23 & 0.00 \\
    $\delta \phi_{T}$&MINERvA 1b+2b & 70.50/23 & 0.00 \\
    $\delta \phi_{T}$& SF & 38.34/23 & 0.02 \\
    $\delta \phi_{T}$& N1p1h & 71.82/23 & 0.00 \\
    \hline
    $\delta \alpha_{T}$&Dipole 1b & 25.14/12 & 0.01 \\
    $\delta \alpha_{T}$&Dipole 1b+2b & 24.15/12 & 0.02 \\
    $\delta \alpha_{T}$&LQCD+MINERvA 1b+2b & 40.39/12 & 0.00 \\
    $\delta \alpha_{T}$&MINERvA 1b+2b & 22.89/12 & 0.03 \\
    $\mathbf{\delta \alpha_{T}}$& \textbf{SF} & $\mathbf{19.99/12}$ & $\mathbf{0.07}$ \\
    $\delta \alpha_{T}$& N1p1h & 24.32/12 & 0.02 \\
    \hline
    \hline
  \end{tabular}
\end{table}
\section{Conclusions}\label{sec:conclusions}

The CCQE model developed in Refs.~\cite{Franco-Munoz23,Franco-Munoz25} for electromagnetic interactions has been extended to the neutrino sector. This framework is based on an unfactorized representation of the spectral function, employing relativistic mean-field momentum distributions and the relativistic distorted-wave impulse approximation. Additionally to the one-body current contribution, it includes two-body meson-exchange currents leading to a 1p1h final state. The model has been implemented in the NEUT event generator and benchmarked against T2K and MINERvA $\nu_{\mu}$ CC0$\pi$ measurements~\cite{T2K20b,T2K18,Lu18,MINERvA20}. Compared to the previous implementation reported in Ref.~\cite{McKean25}, the hadronic tensor is now separated into the one-body vector-vector, vector-axial and axial-axial components, together with the two-body contribution. This decomposition enables the axial form factor to be modified on an event-by-event basis.

Comparisons with the T2K measurements show that, for inclusive data sets, the N1p1h and SF models in NEUT agree better with data, followed by our calculation using the dipole form factor and including only the one-body current contribution. Nevertheless, all considered configurations remain statistically compatible with the data. The TKI variables, which are more sensitive to nuclear effects and hadron kinematics, show a different pattern compared to the inclusive observables.  In this case, the overall agreement deteriorates for all models, indicating that a good description of inclusive observables is not sufficient to ensure agreement with more exclusive measurements. None of our considered configurations is accepted according to the corresponding $p$-values and, among them, the calculation employing the LQCD+MINERvA axial form factor together with two-body meson-exchange currents shows a systematic tendency to overestimate the data and yields the largest $\chi^2$ values. In contrast, comparisons to the MINERvA measurements show that different model configurations can have the best $\chi^{2}$ value depending on the kinematic chosen. Interestingly, for the proton momentum and proton scattering angle, the dipole parametrization with meson-exchange currents is accepted based on the $p$-value. In contrast, for the muon momentum, the dipole parametrization with only one-body currents and the SF model have the better $\chi^{2}$ values.  For the TKI variables, however, all models and configurations are rejected apart from the SF model for the $\delta \alpha_T$ variable.

Overall, the results show that the axial form factor has a significant impact on the predicted cross sections. The LQCD+MINERvA parametrization systematically produces an enhancement of the cross section with respect to the standard dipole form, which in most cases leads to an overestimation of the data.  While the dipole parametrization may provide a better agreement in some cases, it has very limited functional freedom and may therefore not fully capture the true $Q^2$ dependence of the axial form factor. In this context, the MINERvA-only parametrization, which lies between the dipole and the combined LQCD+MINERvA results, provides a more moderate increase of the cross section. For the MINERvA measurements in particular, the level of agreement obtained with this parametrization is comparable to, and in some cases slightly better than, that achieved with the combined LQCD+MINERvA fit, noting that the latter is largely driven by the LQCD parametrization.

The inclusion of two-body meson-exchange currents also leads to an increase in the 1p1h cross sections. A clear preference for this contribution is not observed across all considered measurements, as its inclusion can either improve or worsen the agreement with data depending on the observable and kinematic region. Nevertheless, we remark that the presence of this contribution is well motivated by physical and theoretical considerations. In particular, the sign convention adopted for the $\Delta$-resonance amplitude, which leads to constructive interference with the one-body current contribution, is supported by previous comparisons with electron-nucleus scattering data~\cite{Franco-Munoz23,Franco-Munoz25}, as well as by several theoretical frameworks~\cite{Dekker94,VanderSluys95,Carlson02,Lovato15,Lovato16,Lovato20,Andreoli22,Lovato23,Andreoli24}.

It is important to stress that the comparisons with data discussed in this work do not test the CCQE model in isolation, but the full NEUT generator prediction. Therefore, the reported $\chi^2$ values reflect the combined effect of our CCQE framework together with the other reaction mechanisms contributing to the selected signal, i.e. multi-nucleon excitations and pion production followed by pion absorption in the nuclear medium. Consequently, any agreement or disagreement with data cannot be unambiguously attributed to the CCQE sector alone, since it may also be influenced by the modeling of these additional channels.  In particular, the overestimation observed for the LQCD+MINERvA axial form factor could be tentatively attributed to the combined enhancement arising from the larger axial form factor and the inclusion of meson-exchange currents. However, both ingredients constitute essential components of a consistent description of the reaction and cannot be disentangled from the physical framework. The observed discrepancy may therefore also be influenced by the modeling of the additional reaction mechanisms contributing to the signal. The present comparisons thus reveal general trends and sensitivities, but do not allow for a fully disentangled assessment of the different ingredients entering the simulation.

The results presented in this work pave the way for further applications of the model and contribute to a deeper understanding of lepton-nucleus interactions, moving toward the precision era of neutrino physics.  In the near future, calculations for muon-antineutrino and electron-(anti)neutrino scattering, as well as for other symmetric nuclei such as $^{16}$O, will be implemented in NEUT. Additionally, extending this framework to describe isospin-asymmetric nuclei with partially filled shells, such as $^{40}$Ar, would be especially valuable for experiments such as MicroBooNE~\cite{MicroBooNE20}, SBND~\cite{Szelc16} and DUNE~\cite{DUNE16}.

\begin{acknowledgments}
T. Franco-Munoz  was supported by FWO Junior Postdoctoral Fellowship No. 12AG826N. J.~McKean was supported by Grant-in-Aid for JSPS Research Fellows, JSPS KAKENHI Grant No. 25KF0223. R.G.J. is supported by project RYC2022-035203-I funded by MCIN/AEI/10.13039/501100011033/FEDER and FSE+, UE, and by ``Ayudas para Atracción de Investigadores con Alto Potencial-modalidad A'' funded by VII PPIT-US.
\end{acknowledgments}

\appendix

\section{Total cross section}\label{app:tot_xsec}

For use within NEUT, each hadron tensor table must be accompanied by a total cross section as a function of neutrino energy. This quantity is required by the generator to determine the event rate for a given neutrino flux. In the present implementation, where the axial form factor can be modified through a reweighting of the one-body vector-axial and axial-axial hadron tensor contributions, a separate total cross section table must be generated for each form factor parametrization.

The total cross section is obtained by scanning the incident neutrino energy and, for each energy value, evaluating the sixfold differential cross section over the full sampled phase space. The differential contributions from all generated events are then summed and normalized by the number of thrown events. Finally, this average value is multiplied by the total six-dimensional phase-space volume, yielding the total cross section at that neutrino energy.

\begin{figure}[t]
    \centering
    \includegraphics[width=\linewidth]{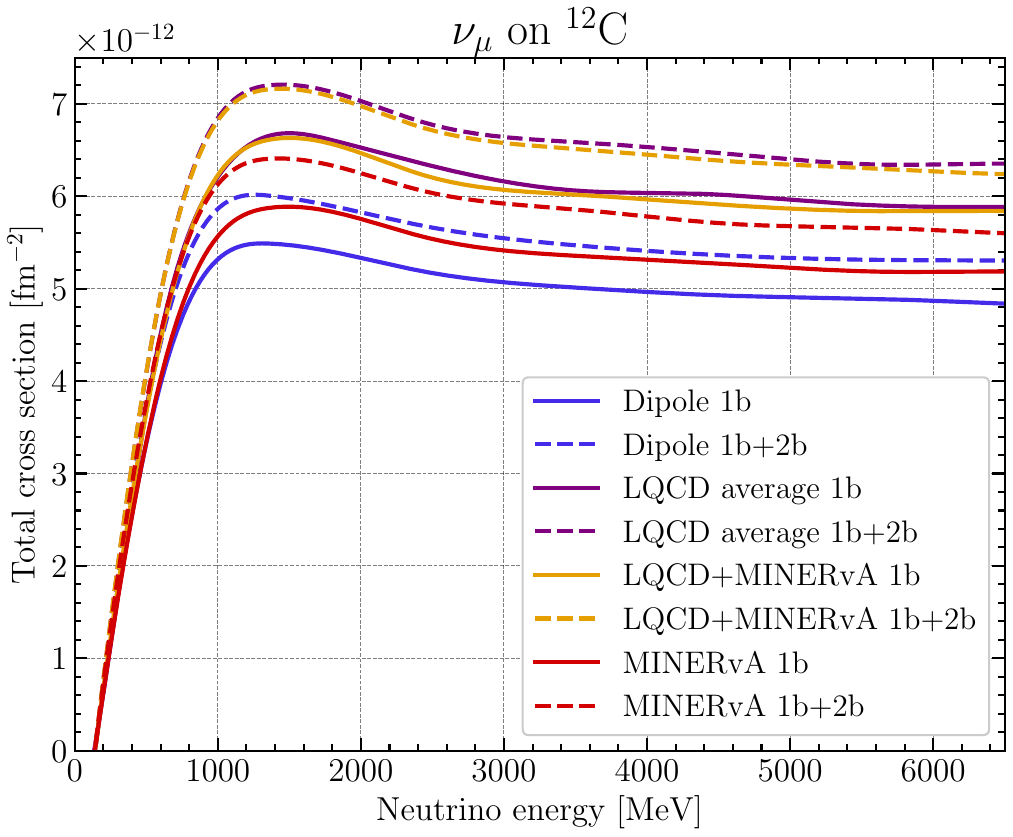}
    \caption{Total CCQE cross section as a function of neutrino energy for $\nu_\mu$ scattering on $^{12}$C. Results are shown for the different axial form factor parametrizations considered in this work (standard dipole with $M_A = 1.05$ GeV, MINERvA fit~\cite{MINERvA23}, LQCD+MINERvA fit~\cite{Meyer25} and the LQCD average result from~\cite{Meyer26} is also shown for reference), both with and without the contribution from two-body meson-exchange currents.}
    \label{fig:tot_xsec_appendix}
\end{figure}

Figure~\ref{fig:tot_xsec_appendix} shows the resulting total cross sections as a function of neutrino energy for the different axial form factor parametrizations considered in this work, with and without including the contribution from two-body meson-exchange currents. As expected, larger values of $G_A(Q^2)$, as well as the inclusion of two-body currents, lead to an overall enhancement of the total cross section.

\bibliography{bibliography}

\end{document}